\documentclass[final]{elsarticle}

\usepackage{lineno,hyperref}
\usepackage{tabularx}
\usepackage{caption}
\usepackage{amsmath,amssymb,amsfonts}
\usepackage{algorithmic}
\usepackage{float}
\usepackage{placeins}
\usepackage{graphicx}
\usepackage{textcomp}
\usepackage{multicol}
\usepackage{multirow}
\usepackage{commath}
\usepackage{blindtext}
\usepackage{subcaption}
\usepackage{balance}
\usepackage{array,multirow}
\usepackage{natbib}
\usepackage{physics}
\usepackage{tabularx}
\usepackage{adjustbox}

\journal{Pattern Recognition}

\begin{document}

\begin{frontmatter}
\title{Time accelerated image super-resolution using shallow residual feature representative network}

\author[1]{Meenu Ajith\corref{cor1}}%
\ead{majith@unm.edu}
\author[1]{Aswathy Rajendra Kurup}

\author[1]{Manel Mart\'{i}nez-Ram\'on} 

\cortext[cor1]{Corresponding author}
\address[1]{Department of Electrical and Computer Engineering, The University of New Mexico, New Mexico, 87106, USA}

\begin{abstract}
The recent advances in deep learning indicate significant progress in the field of single image super-resolution. With the advent of these techniques, high-resolution image with high peak signal to noise ratio (PSNR) and excellent perceptual quality can be reconstructed. The major challenges associated with existing deep convolutional neural networks are their computational complexity and time; the increasing depth of the networks, often result in high space complexity. To alleviate these issues, we developed an innovative shallow residual feature representative network (SRFRN) that uses a bicubic interpolated low-resolution image as input and residual representative units (RFR) which include serially stacked residual non-linear convolutions. Furthermore, the reconstruction of the high-resolution image is done by combining the output of the RFR units and the residual output from the bicubic interpolated LR image. Finally, multiple experiments have been performed on the benchmark datasets and the proposed model illustrates superior performance for higher scales. Besides, this model also exhibits faster execution time compared to all the existing approaches.
\end{abstract}

\begin{keyword}
Super-resolution\sep Convolutional neural networks\sep Residual learning 
\end{keyword}

\end{frontmatter}

\section{Introduction}

Super-resolution (SR) is the process of enhancement of the resolution of an image where the high resolution (HR) image is estimated from its low resolution (LR) counterpart \cite{ledig2017photo}. SR finds its application in many real-time problems from different fields such as microbiology \cite{fornasiero2015super}, medical imaging \cite{robinson2017new}, satellite imaging \cite{genitha2010super,zhang2014super} and in many Computer vision applications \cite{yeh2016semantic,nasrollahi2014super}. 

SR techniques can be classified into two broad classes based on the number of input images: Single input super-resolution (SISR) and multi-input super-resolution (MISR) \cite{yang2019deep}. SISR has an advantage over MISR in terms of its lower computational burden. Compared to the MISR techniques, which can improve the actual resolution by combining the information from different images, SISR techniques improve the image only perceptually, but this is often what the user needs. SISR techniques include algorithms that can be divided into three categories: methods based on interpolation such as bicubic interpolation \cite{keys1981cubic}, reconstruction based techniques \cite{douchon1979lanczos,sun2008image, yan2015single} and learning-based methods that make use of machine learning techniques to learn from the training examples. Markov Random Field (MRF) \cite{freeman2002example}, Neighbor embedding method, Manifold learning \cite{chang2004super}, sparse coding, Anchored regression \cite{salvador2016example} are a few learning-based techniques. Interpolation based methods are computationally inexpensive compared to the other two methods but they tend to show lower accuracy. Reconstruction based techniques can be time-consuming and their performance drops with the increase in the scale factor. Learning-based methods are known for their efficiency in both computation and performance \cite{yang2019deep}. Deep learning-based approaches are becoming more popular in this field due to their outstanding performance compared to other approaches. The traditional SR approaches are unable to remove the defects and artifacts due to compression while upscaling. Deep learning methods can solve this problem to a great extent \cite{yang2019deep}. Deep learning techniques have the advantage over other techniques due to their capability to extract high-level abstractions that help in getting a better mapping from LR space to HR space. Recently developed Convolutional neural network (CNN) and Generative adversarial network-based techniques have shown outstanding performance compared to the traditional approaches \cite{ledig2017photo} but the main drawback of these approaches is their large training time and computational cost which makes these techniques hard to be implemented during a real-time scenario.

The main contributions of this paper are five-fold:
\begin{itemize}
\item 	We introduce a novel structure which stores residual information from the previous layers instead of the features themselves. The residual learning allows the network to avoid redundancies.
\item The proposed shallow residual feature representative network (SRFRN) uses depthless convolutions resulting in a lightweight network. Hence in terms of training and testing time the network exhibits superior performance compared to the state-of-the-art networks.
\item A patch extraction technique was used as pre-processing before passing the LR image into the network. This further reduced the time complexity of the model.
\item In this work we introduce residual feature representative  (RFR) units, which consist of convolutional blocks that help to extract features and store differential information.
\item Finally, the network shows less degradation for higher scales. Therefore, different experiments conducted on various databases with $\times$4 scale have shown outstanding performance compared to the existing approaches.
\end{itemize}

Different experiments have been conducted in order to measure the performance of our approach. Results show a significant advantage of the introduced methodology in terms of PSNR and computational time.

\section{Related work}
Over the years, several deep learning approaches have been implemented to address the SR problem. These approaches were mainly based on CNNs, of which, the first model was implemented by Dong et al. named Super-Resolution Convolutional Neural Network (SRCNN) which was inspired by the conventional Sparse-coding based SR methods  \cite{dong2014learning}. The structure of SRCNN is fully feed-forward with learning upscaling filters and requires little pre and post-processing other than the optimization. SRCNN has the advantage of the simplicity of its structure \cite{dong2014learning} and good performance but the training time for this structure is very large. Further, in 2016, an accelerated version of SRCNN was introduced by Dong et al. named Fast Super-Resolution Convolutional Neural Networks (FSRCNN)\cite{dong2016accelerating}. This structure was implemented to address the time and computational complexity of SRCNN. FSRCNN was the first structure to use a deconvolutional layer for the reconstruction of HR from LR features. The structure was able to achieve a  speed 40 times higher, superior accuracy and comparable image restoration quality to that of SRCNN \cite{dong2016accelerating}. The deconvolution layer makes use of nearest-neighbor interpolation while reconstructing which causes repetition of unsampled features in each direction \cite{yang2019deep}. Efficient sub-pixel convolution neural network (ESPCN) by Shi et al. tackles this problem by simplifying the deconvolution layer into sub-pixel convolution.\cite{yang2019deep,shi2016real}. 

Apart from the smaller networks, several very deep networks were also used in SISR. One of the very first deep models was VDSR \cite{kim2016accurate} which is a 20 weight layered VGG-net \cite{veit2016residual}. VDSR takes the bicubic interpolated image as input and the network learns to map from this interpolated image to the differential features between HR and bicubic. During training, different scale factors of the bicubic interpolated image are assembled. In 2016, Kim et al. \cite{Kim_2016_CVPR} introduced the Deep recursive convolutional network (DRCN) that uses the same convolutional layer repeatedly as a result of which the number of parameters does not increase. The network uses the feature maps from each recursion to obtain a reconstructed high-resolution image (HR). The HR predictions from different levels of recursions were further combined to achieve a more accurate prediction. While fusing these HR outputs the scalar weights are not adaptive, that is, the scalar weights don't change with different inputs. DRCN and VDSR show similar performance except that DRCN uses a multi-supervised strategy for training. Due to their deep structure, they have a greater training time compared to other shallow structured networks. Eventually, the idea of residual units was incorporated into several networks. Residual units, in general, comprises of non-linear convolutions and residual learning. SRResNet \cite{ledig2017photo} inspired from ResNet \cite{he2016identity} was the first network to implement these residual units and it focused on reconstructing photo-realistic textures from LR images. Later on, the Deep Recursive residual network (DRRN) was proposed by Tai et al. which used the technique of residual learning and recursive blocks to control model parameters and to tackle training difficulty \cite{8099781}. Residual dense network (RDN)\cite{zhang2018residual} proposed by Zhang et al. make use of hierarchical features from the convolutional layers to perform SR. The basic units used here are the densely connected Residual dense blocks (RDB). The main function of RDB is to extract dense local features. The local features are further combined globally using dense connections and residual learning.   Progressive property of ResNet has been adopted in the DEGREE \cite{yang2017deep} which combined it with the sub-band reconstruction technique used in traditional methods. The high-frequency details were reconstructed using a recursive residual block. Another such network that makes use of this progressive property is the Laplacian Pyramid Super-Resolution Network (LapSRN) \cite{lai2017deep}. These networks don't use bicubic interpolation as the initial step and the progressive reconstruction is applied to multi-scale HR predictions. The deeper nature of the structure poses training difficulties in terms of a large number of parameters, especially for larger scales.

\section{Proposed Approach}

\begin{figure*}[t]
  \centering
  \captionsetup{justification=centering}
  \includegraphics[width=1\textwidth]{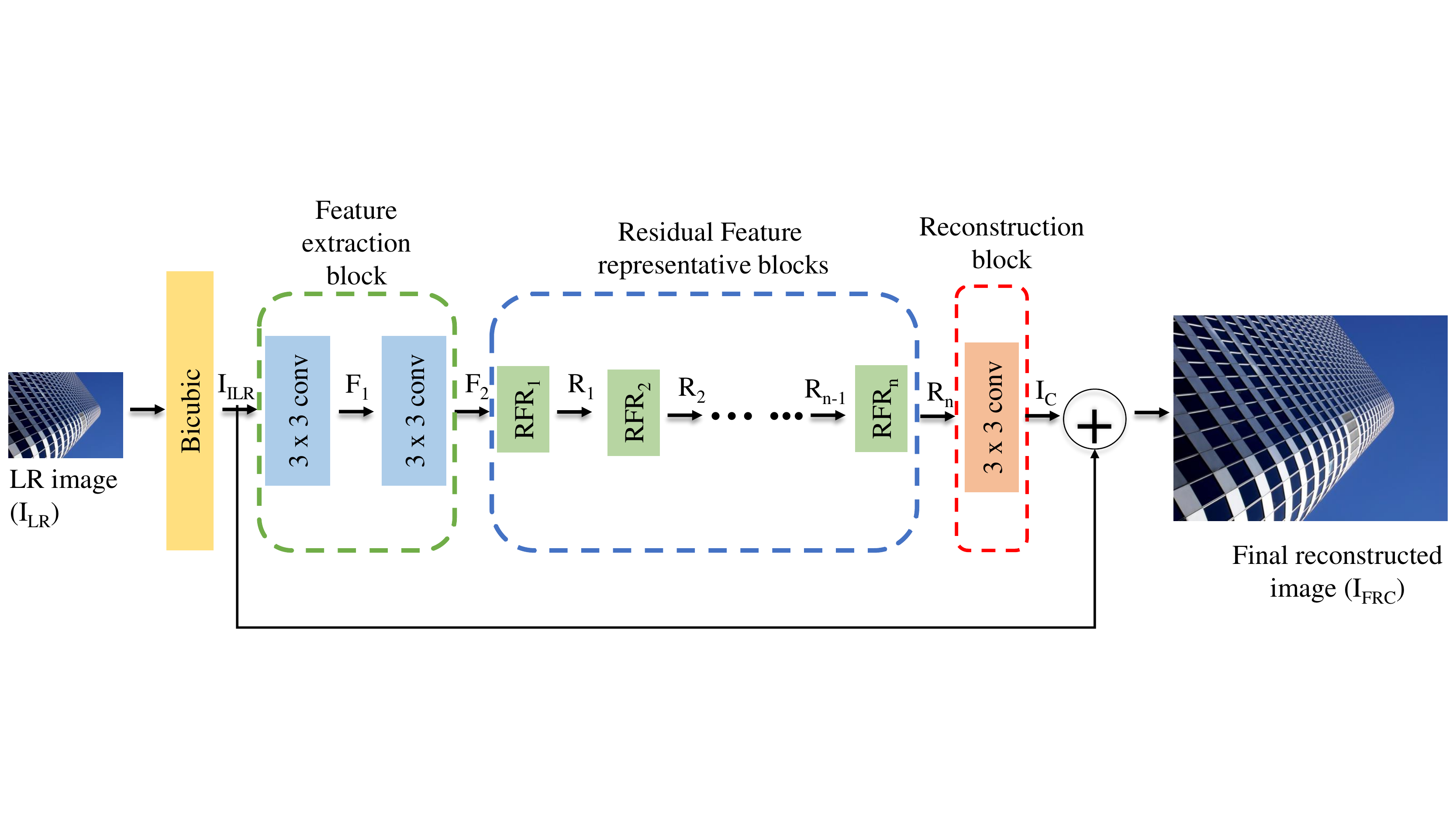}
    \caption{The architecture of the proposed approach.}
   \label{fig1: p1}
 \end{figure*}
 
The structure of SRFRN consists of four parts as shown in Fig \ref {fig1: p1}: bicubic interpolation block, feature extraction blocks, residual feature representative blocks, and a reconstruction block. Let $I_{LR}$ and $I_{HR}$ denote the input LR image and the ground truth HR image. Initially, the bicubic block is used to upsample the input LR image and the resulting image is passed onto the next block. \par
\begin{equation}
    I_{ILR}=B(I_{LR})
\end{equation}
where $I_{ILR}$ denotes the interpolated LR image and $B(\cdot)$ represents the bicubic interpolation operation.
This predefined up sampling helps in the modeling of image details and creates an interpolated image of the required size. The definitive knowledge of the domain is highly beneficial for the following convolutional blocks. Moreover, similar structures have been used in several super-resolution structures \cite {Timofte_2013_ICCV, BMVC.26.135}. Next, the feature extraction block consists of two convolutional layers with $64$ filters each and a kernel size of $3\times3$. This block outputs the feature maps of the upsampled image. The convolutional layers that extract the features can be formulated as follows:\par
\begin{equation}
F_{1} =W_{1}\ast I_{ILR}+B_{1}
\end{equation}

\begin{equation}
F_{2}=W_{2}\ast F_{1} +B_{2}
\end{equation}
where $\ast$ is the convolution operator.

Here $F_{1}$ and $F_{2}$ represents the features extracted from the first and second layers of the feature extraction block. Further, the filters are denoted as $W_{1}$ and $W_{2}$ and the respective biases are $B_{1}$ and $B_{2}$ 
The feature extraction block is followed by a series of residual feature representative blocks (RFR). The network is capable of accommodating $n$ RFR units and each of them consists of 3 convolutional blocks that use a kernel of size $3\times3$. The output from the $n^{\text{th}}$ RFR block can be represented as follows:
\begin{equation}
R_{n}=f_{RFR_{n}}(R_{n-1})
\end{equation}
where $f_{RFR_{n}}(\cdot)$ is a function consisting of convolutions and leaky rectified linear units \cite{Maas13rectifiernonlinearities}.
The RFR units are designed to work based on residual learning. This technique was employed to enhance the memory of the network and to combine information from the former convolutional blocks. \par
\begin{figure*}[t]
  \centering
  \captionsetup{justification=centering}
  \includegraphics[scale=0.35]{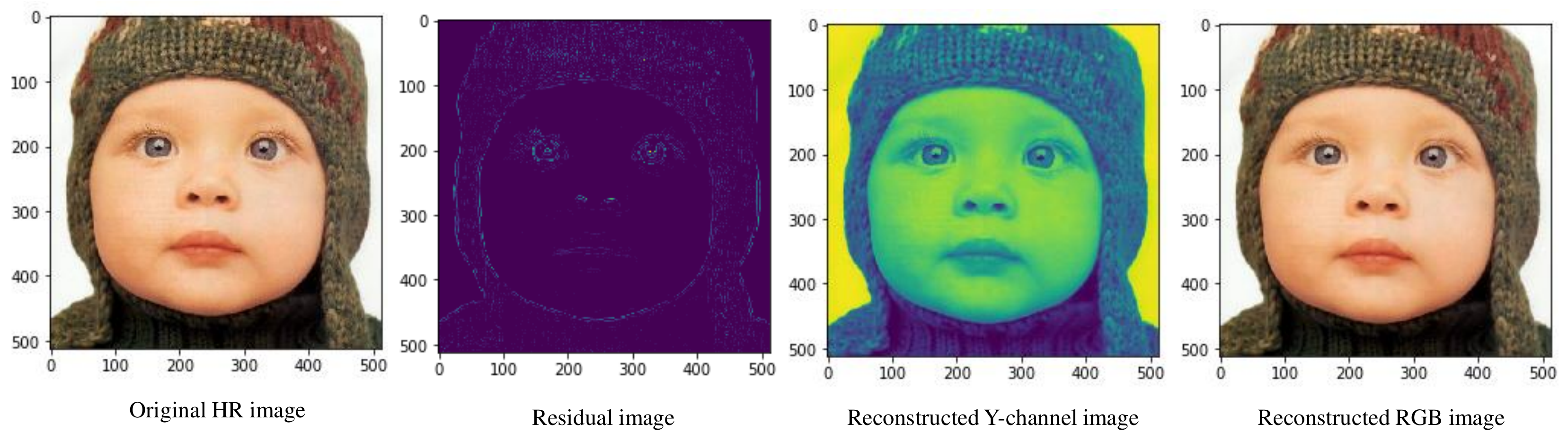}
    \caption{The original HR image and other image representations extracted from the model.}
   \label{fig3:p3}
 \end{figure*}
Finally, in the reconstruction block, another $3\times3$ convolution is performed on the hierarchical features extracted from the RFR units. A feature fusion method is employed as the last step and the output from the reconstruction layer is concatenated with the output of the bicubic interpolation block from the initial stage. The predicted HR image from the proposed model can be expressed as follows:
\begin{equation}
    I_{FRC}=I_{ILR}+I_{C}
\end{equation}
where $I_{FRC}$ and $I_{C}$ represents the final reconstructed output and the convolutional output.\par

\begin{figure*}[t]
  \centering
  \captionsetup{justification=centering}
  \includegraphics[width=1\textwidth]{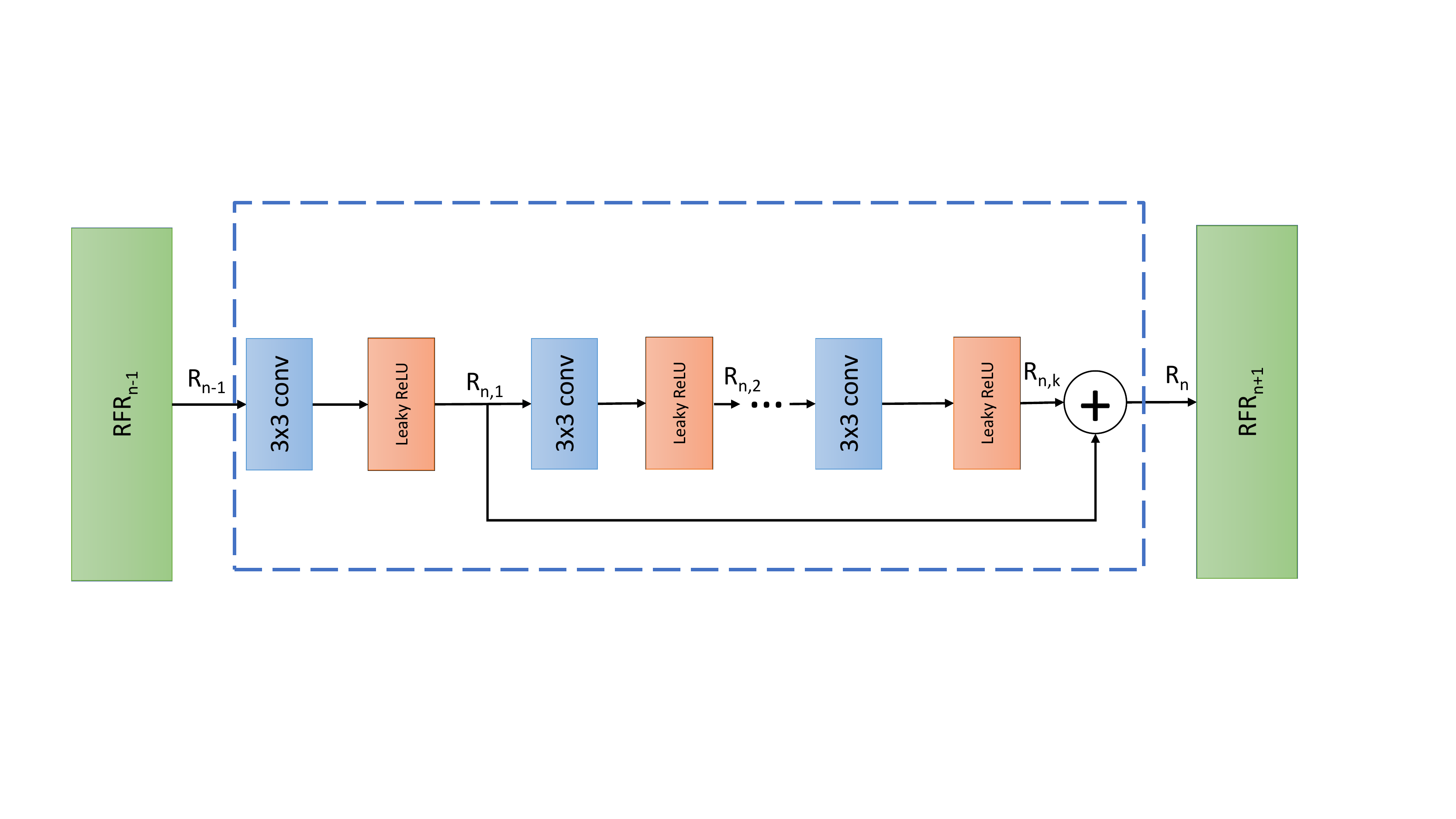}
    \caption{Residual feature representative blocks}
   \label{fig2: p2}
 \end{figure*}
\subsection{Residual Representative Block}
The residual learning-based CNN has been proven to more powerful than the non-residual networks in several works related to SR \cite{Lim2017EnhancedDR,10.5555/3298023.3298188, Kim2015AccurateIS, 599c796c601a182cd263bb8a}. The RFR blocks used in this work consist of a series of shallow convolutional layers stacked together.
Additionally, each of the convolutional layers is followed by a Leaky ReLU activation function. The Leaky ReLU is used over ReLU since the latter gets saturated at the negative area and hence the gradient becomes zero. If this happens, the training stalls in these neurons during the backpropagation step. 
The output of the $n^\text{th}$ RFR block can be expanded and formulated as:
\begin{equation}
    R_{n}=R_{n,k}+R_{n,1}
\end{equation}
\begin{equation}
    R_{n,1} =\max \left\lbrace 0.1(W_{n,1}\ast R_{(n-1),1}+B_{n,1}),(W_{n,1}\ast R_{(n-1),1}+B_{n,1})\right\rbrace
\end{equation}
\begin{equation}
    R_{n,k}=\max\left\lbrace(0.1(W_{n,k}\ast R_{(n-1),(k-1)}+B_{n,k}),(W_{n,k}\ast R_{(n-1),(k-1)}+B_{n,k})\right\rbrace
\end{equation}
where $R_{n,k}$ and $R_{n,1}$ denote the output of the $1_{st}$ and $k_{th}$ convolutional layers of the given RFR block. Here $R_{n}$ also represents the feature fusion at the end of each RFR block. This technique helps in significant improvement in information flow. Finally, it also acts as a local error-correcting feedback system to accelerate the training and thereby resulting in faster convergence. Since the basic blocks of RFR consist of standard convolutional units that are not deep, this structure is easily trainable, and its depth can be reduced without compromising the PSNR. The detailed structure is further illustrated in Fig. \ref {fig2: p2}

\subsection{Loss Function}
In the super-resolution problem, several loss functions were introduced to minimize the reconstruction error of the generated HR image. The most common amongst them is the pixel-wise L1 and L2 loss. The sum of all the absolute errors between each pixel in the reconstructed and ground truth HR image is calculated to obtain the L1 loss. Meanwhile, the L2 loss was unable to provide high-quality reconstruction images and often lead to smooth and blurry textures in them. Further, the convergence rate of L1 loss was much faster compared to the L2 loss. Thus, minimizing the L1 loss results in a relatively high peak signal to noise ratio (PSNR) with less complexity. But the main disadvantage of the pixel-wise loss function is that it does not improve the visual perception of the generated image. In order to alleviate this issue, several other loss functions such as content loss \cite{Johnson2016Perceptual}, adversarial loss \cite{ NIPS2014_5423} and texture loss \cite{ NIPS2015_5633} were introduced. Though they succeeded in improving the perceptual quality, the computational complexity was comparatively high. Since the PSNR improvement was not significant in our experiments, in this work we only consider the L1 loss, which is relatively less complex. Hence the optimization of the proposed model is done by minimizing the difference between the ground truth $I_{HR}$ and the predicted image $I_{FRC}$. This error minimization is formulated as follows:
\begin{equation}
    L1=\frac{1}{N}\sum_{i=1}^{N}\norm{I_{FRC}-I_{HR}}_{1}
\end{equation}

\section{Experiments}
\subsection{Databases}
The training of this model was conducted on two datasets and it utilized 200 images from the Berkeley Segmentation Dataset (BSD) \cite{Martin2001ADO} and 91 images from Yang dataset \cite{10.1109/TIP.2010.2050625}. Data augmentation methods such as rotation, flipping, and scaling were done to increase the training data to 2328 images. Further, the testing was done on four widely used datasets: Set5 \cite{bevilacqua:hal-00747054}, Set14 \cite{10.1007/978-3-642-27413-8_47}, BSD100 \cite{DBLP:conf/accv/TimofteSG14} and Urban100 \cite{ad268b48b19d4cdbac2d68d1934ef44e}. These databases include various natural and urban scenes from diverse environments. Here the performance evaluation with peak signal-to-noise ratio (PSNR) and structural similarity (SSIM) \cite{1284395} is done only on the luminance channel (Y channel) of the YCbCr color space.
\subsection{Training}
The proposed SRFRN model was trained on 2328 images with a 6-layer deep convolutional architecture. The different scaling factors considered are 2, 3 and 4 and the downsampling was done using bicubic interpolation.  To accelerate the training process, a patch extraction technique was employed as a preprocessing technique. At first, smaller patches of sub-images were extracted from the given HR images. A patch size of $m\times n$, much smaller than the minimum image size in the dataset was utilized to minimize the computational burden. These patches were further downsampled to create the corresponding LR sub-images. Hence the entire training was conducted using these sub-images with a batch size of 24. The learning rate for training is originally fixed to $10^{-3}$ and is decreased if the validation error does not change for successive 10 epochs. The model uses the Adam optimizer \cite{ Kingma2014AdamAM} and the loss function for optimizing this network is the mean absolute error or L1. The proposed model is trained for 50 iterations to obtain the optimal parameters and the total training time is 20 minutes on an NVIDIA GeForce GTX 1060 GDDR5 GPU.

\section{Results}

\begin{table}[]
  \begin{adjustbox}{max width=\textwidth}
\begin{tabular}{|l|l|l|l|l|l|}
\hline
Scale & Model   & \begin{tabular}[c]{@{}l@{}}Set5\\ PSNR/SSIM\end{tabular} & \begin{tabular}[c]{@{}l@{}}Set14\\ PSNR/SSIM\end{tabular} & \begin{tabular}[c]{@{}l@{}}BSD100\\ PSNR/SSIM\end{tabular} & \begin{tabular}[c]{@{}l@{}}Urban100\\ PSNR/SSIM\end{tabular} \\ \hline
      & Bicubic & 33.66/0.9299                                             & 30.24/0.8688                                              & 29.56/0.8431                                               & 26.88/0.8403                                                 \\ \cline{2-6} 
      & SRCNN   & 36.66/0.9542                                             & 32.42/0.9063                                              & 31.36/0.8879                                               & 29.50/0.8946                                                 \\ \cline{2-6} 
      & FSRCNN  & 36.99/0.9550                                             & 32.73/0.909                                               & 31.51/0.8910                                               & 29.87/0.9010                                                 \\ \cline{2-6} 
2     & VDSR    & 37.53/0.9587                                             & 33.03/0.9124                                              & 31.90/0.8960                                               & 30.76/0.9140                                                 \\ \cline{2-6} 
      & LapSRN  & 37.52/0.9592                                             & 33.08/0.9132                                              & 31.80/0.8958                                               & 30.41/0.9101                                                 \\ \cline{2-6} 
      & RDN     & \textbf{38.24}/0.9614                                            & 34.01/0.9212                                              & 32.34/0.9017                                               & \textbf{32.89}/0.9353                                                 \\ \cline{2-6} 
      & SRFRN (ours)    & 36.58/\textbf{0.9635}                                            & \textbf{34.18/0.9317}                                              & \textbf{33.57/0.9133}                                             & 32.67/\textbf{0.9337}                                                \\ \hline
      & Bicubic & 30.39/0.8682                                             & 27.55/0.7742                                              & 27.21/0.7385                                               & 24.46/0.7349                                                 \\ \cline{2-6} 
      & SRCNN   & 32.75/0.9090                                             & 29.28/0.8209                                              & 28.41/0.7863                                               & 26.44/0.8088                                                 \\ \cline{2-6} 
      & FSRCNN  & 33.16/0.9140                                             & 29.42/0.8242                                              & 28.52/0.7893                                               & 26.41/0.8064                                                 \\ \cline{2-6} 
3     & VDSR    & 33.66/0.9213                                             & 29.77/0.8314                                              & 28.82/0.7976                                               & 27.14/0.8279                                                 \\ \cline{2-6} 
      & LapSRN  & 33.78/0.9214                                             & 29.87/0.8333                                              & 28.81/0.7970                                                & 27.06/0.8270                                                  \\ \cline{2-6} 
      & RDN     & \textbf{34.71}/0.9296                                             & 30.57/0.8468                                              & 29.26/0.8093                                               & 28.80/0.8653                                                 \\ \cline{2-6} 
      & SRFRN (ours)    & 34.30/\textbf{0.9336}                                             & \textbf{32.77/0.8552}                                             & \textbf{32.38/0.8226}                                              & \textbf{31.40/0.8669}                                                \\ \hline
      & Bicubic & 28.42/0.8104                                             & 26.00/0.7027                                              & 25.96/0.6675                                               & 23.14/0.6577                                                 \\ \cline{2-6} 
      & SRCNN   & 30.48/0.8628                                             & 27.49/0.7503                                              & 26.90/0.7101                                               & 24.79/0.7374                                                 \\ \cline{2-6} 
      & FSRCNN  & 30.71/0.8650                                              & 27.70/0.7560                                              & 26.97/0.7140                                               & 24.61/0.7271                                                 \\ \cline{2-6} 
4     & VDSR    & 31.35/0.8838                                             & 28.01/0.7674                                              & 27.29/0.7251                                               & 25.18/0.7524                                                 \\ \cline{2-6} 
      & LapSRN  & 31.54/0.8850                                             & 28.19/0.7720                                              & 27.32/0.7281                                               & 25.21/0.7563                                                 \\ \cline{2-6} 
      & RDN     & 32.47/0.8990                                             & 28.81/0.7871                                              & 27.72/0.7419                                               & 26.61/\textbf{0.8028}                                                 \\ \cline{2-6} 
      & SRFRN (ours)    & \textbf{33.14/0.9137}                                             & \textbf{31.95/0.7931}                                              & \textbf{31.77/0.7522}                                               & \textbf{30.74}/0.7998                                                \\ \hline
\end{tabular}
\end{adjustbox}
\caption{Comparison of average PSNR(dB) and SSIM among the CNN-based methods and bicubic interpolation on test datasets Set5, Set14, B100 , and Urban100 for scale
factor $\times$2, $\times$3, $\times$4.}
    \label{fig:Table1}
\end{table}

\begin{figure}
    \centering
    \includegraphics[scale=0.6]{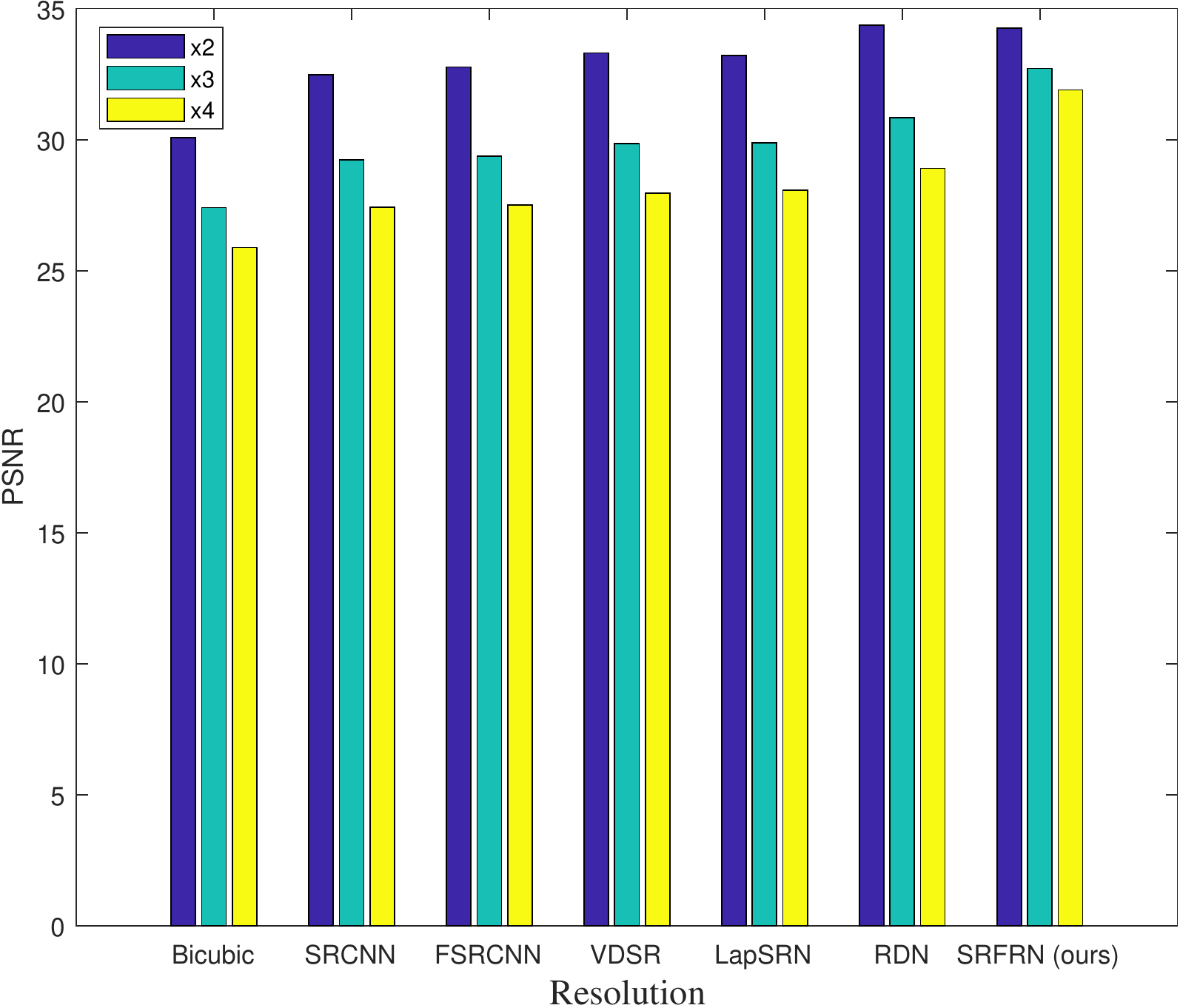}
    \caption{Comparison of the average PSNR of all methods for different resolutions. }
    \label{fig4:p4}
\end{figure}

\begin{figure}[!htb]
\begin{tabularx}{\linewidth}{XX}
\begin{minipage}{\linewidth}
     \centering
     \includegraphics[width=0.6\linewidth,height=3cm]{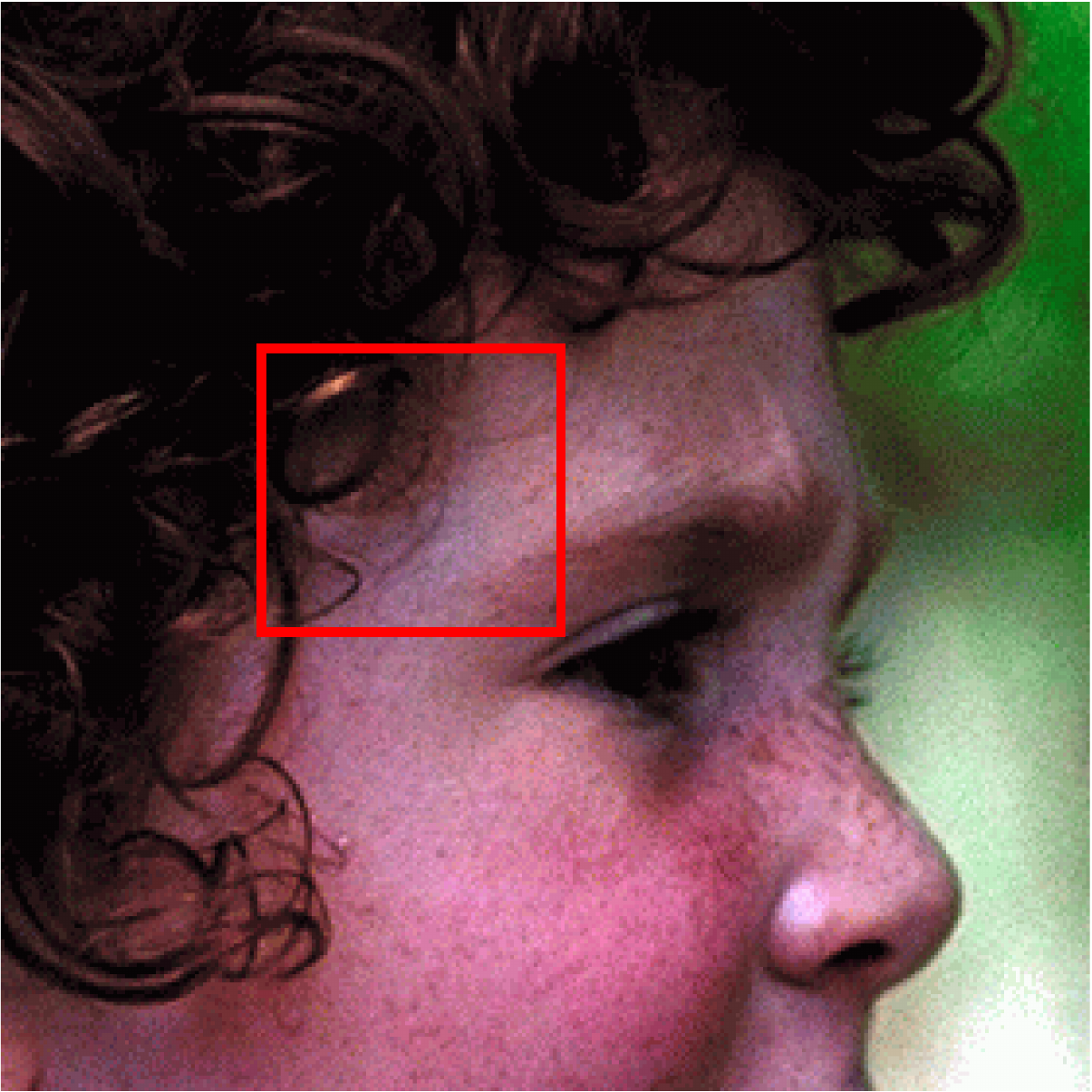}
     \subcaption{}\label{Fig:Data0}
    \end{minipage}
   &
   \begin{minipage}{\linewidth}
   \begin{minipage}{0.25\linewidth}
     \centering
     \includegraphics[scale=0.2]{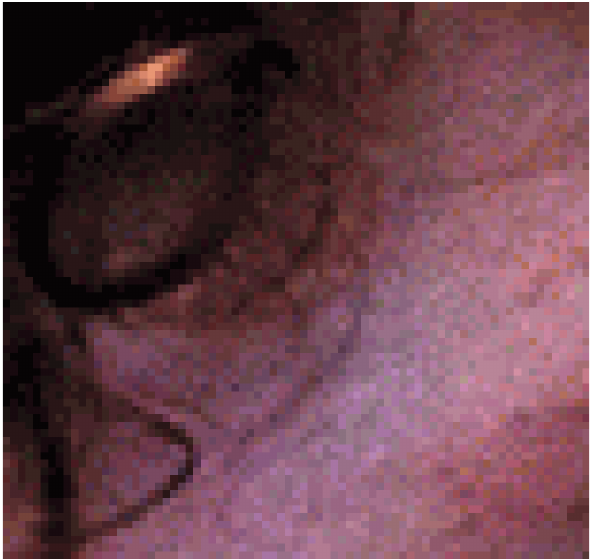}
     \subcaption{}\label{Fig:Data1}
   \end{minipage}\hfill
   \begin{minipage}{0.25\linewidth}
     \centering
     \includegraphics[scale=0.2]{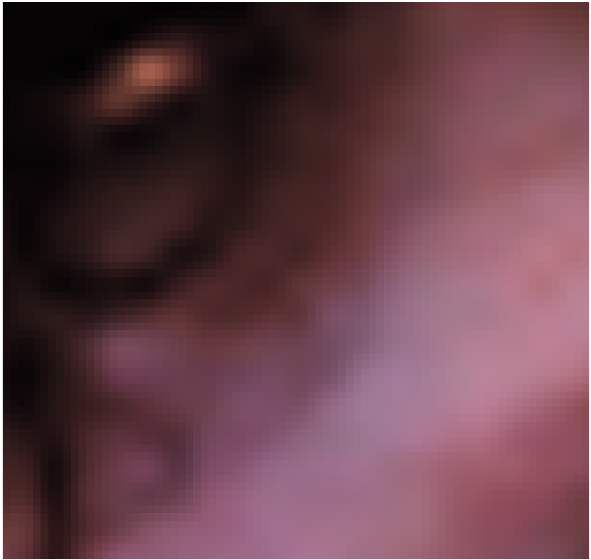}
     \subcaption{}\label{Fig:Data2}
   \end{minipage}\hfill
    \begin{minipage}{0.25\linewidth}
     \centering
     \includegraphics[scale=0.2]{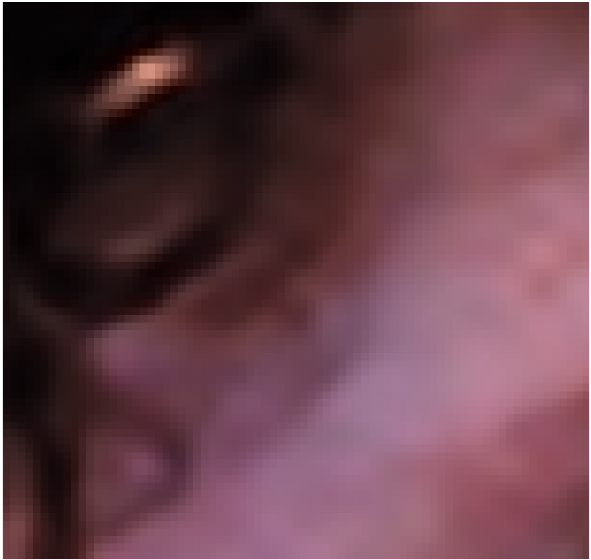}
     \subcaption{}\label{Fig:Data3}
   \end{minipage}\hfill
   \begin{minipage}{0.25\linewidth}
     \centering
     \includegraphics[scale=0.2]{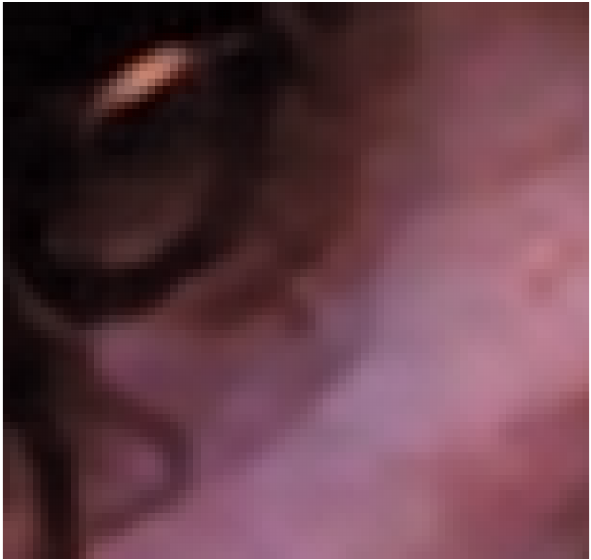}
     \subcaption{}\label{Fig:Data4}
   \end{minipage}

\begin{minipage}{0.25\linewidth}
     \centering
     \includegraphics[scale=0.2]{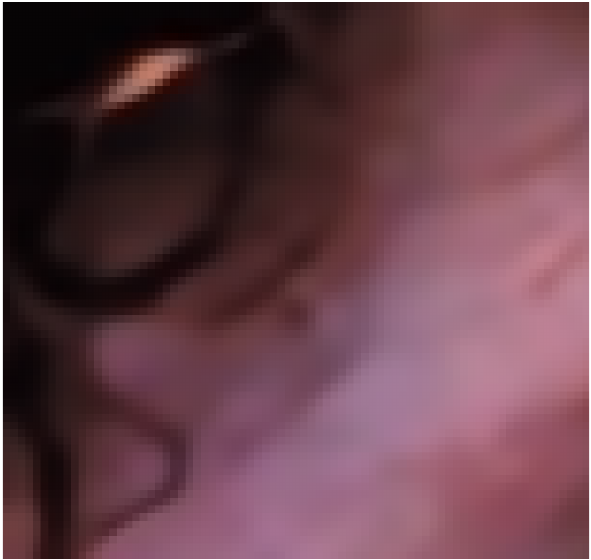}
     \subcaption{}\label{Fig:Data5}
   \end{minipage}\hfill
   \begin{minipage}{0.25\linewidth}
     \centering
     \includegraphics[scale=0.2]{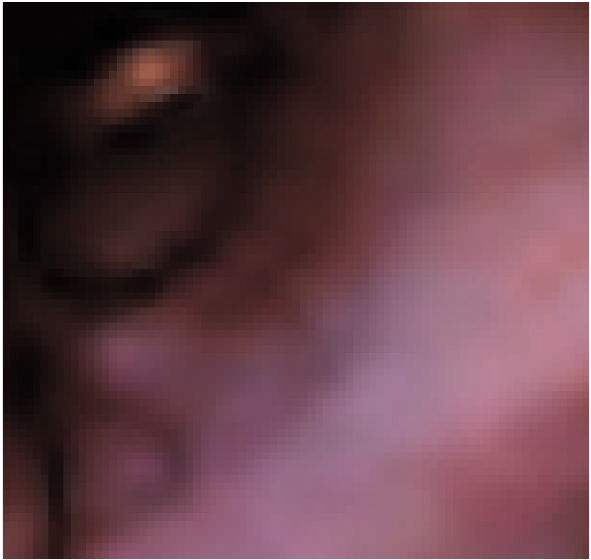}
     \subcaption{}\label{Fig:Data6}
   \end{minipage}\hfill
    \begin{minipage}{0.25\linewidth}
     \centering
     \includegraphics[scale=0.2]{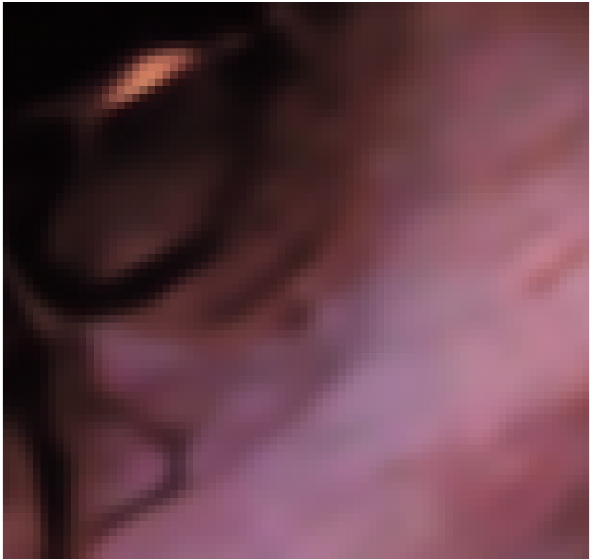}
     \subcaption{}\label{Fig:Data7}
   \end{minipage}\hfill
   \begin{minipage}{0.25\linewidth}
     \centering
     \includegraphics[scale=0.2]{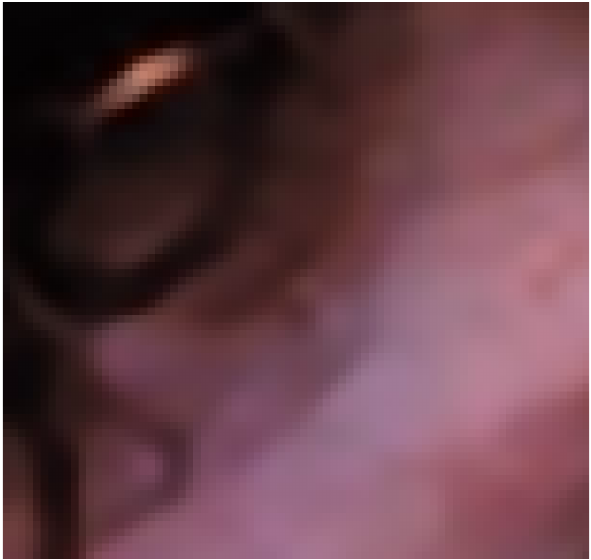}
     \subcaption{}\label{Fig:Data8}
   \end{minipage}
     \end{minipage}
    \end{tabularx}
\caption{Image super resolution with scale $\times$3. (a) Set5:img$\_004$, (b) HR, (c) Bicubic, (d) SRCNN, (e) FSRCNN, (f) LapSRN, (g) VDSR, (h) RDN and (i) SRFRN (ours). }\label{fig5:p5}
\end{figure}

\begin{figure}[!htb]
\begin{tabularx}{\linewidth}{XX}
\begin{minipage}{\linewidth}
     \centering
     \includegraphics[width=0.7\linewidth,height=3.5cm]{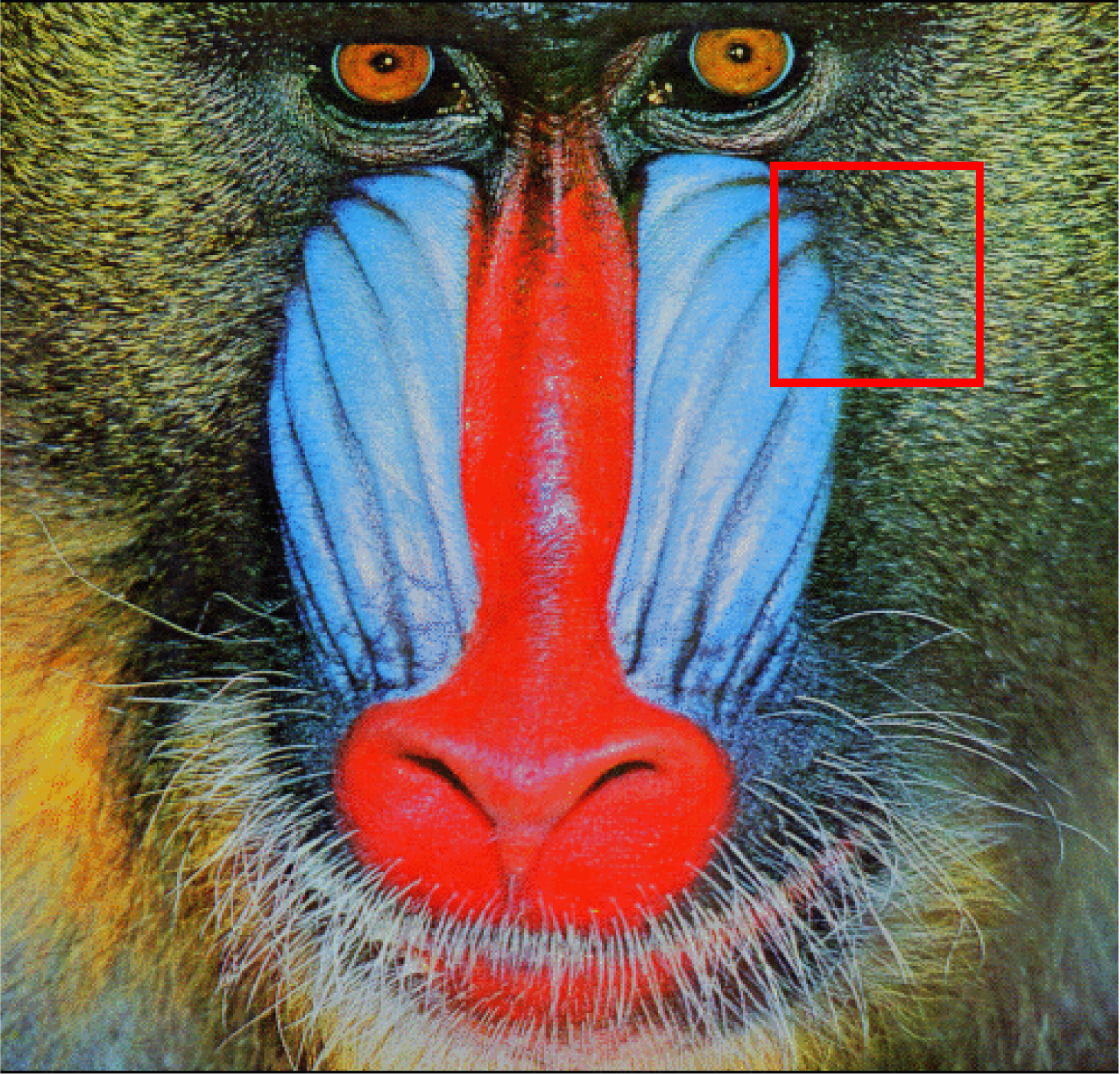}
     \subcaption{}\label{Fig:Data01}
    \end{minipage}
   &
   \begin{minipage}{\linewidth}
   \begin{minipage}{0.25\linewidth}
     \centering
     \includegraphics[scale=0.2]{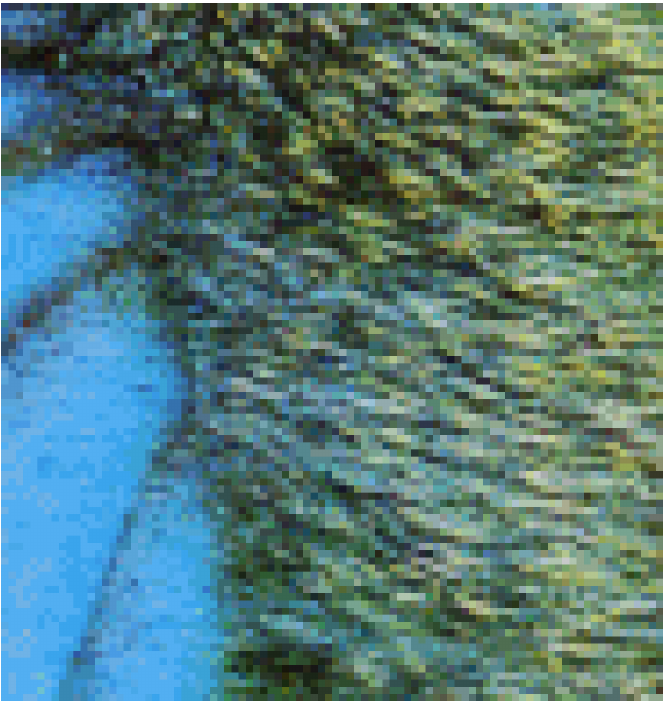}
     \subcaption{}\label{Fig:Data11}
   \end{minipage}\hfill
   \begin{minipage}{0.25\linewidth}
     \centering
     \includegraphics[scale=0.2]{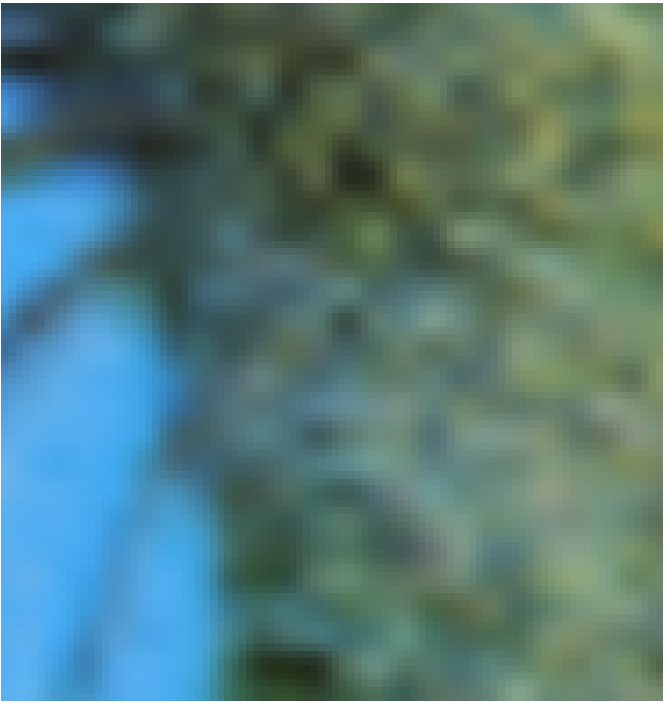}
     \subcaption{}\label{Fig:Data21}
   \end{minipage}\hfill
    \begin{minipage}{0.25\linewidth}
     \centering
     \includegraphics[scale=0.2]{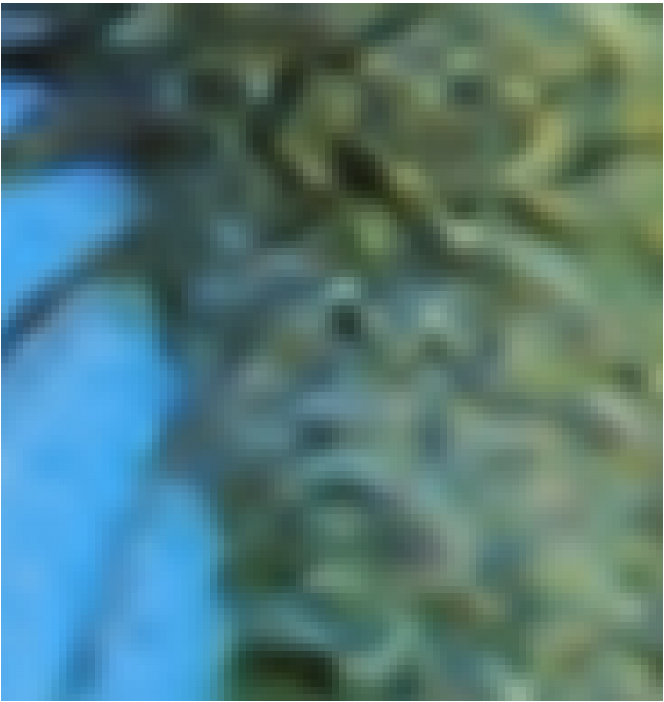}
     \subcaption{}\label{Fig:Data31}
   \end{minipage}\hfill
   \begin{minipage}{0.25\linewidth}
     \centering
     \includegraphics[scale=0.2]{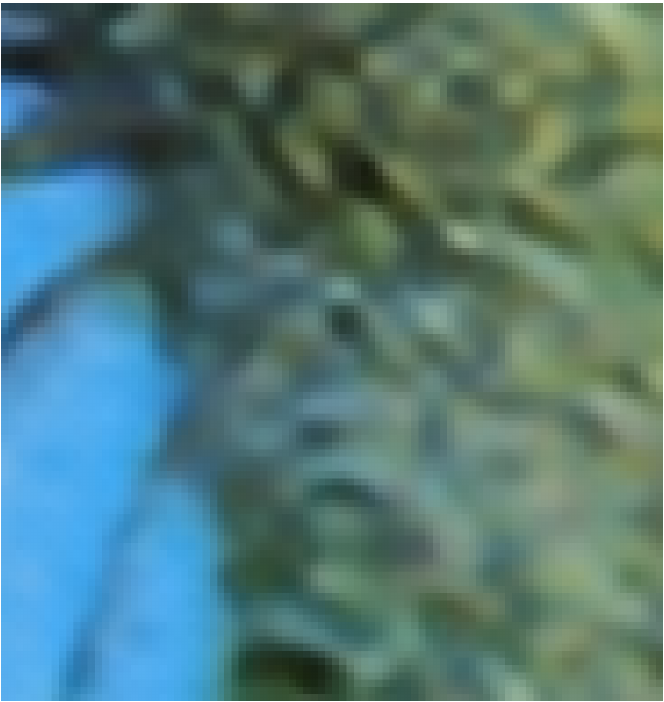}
     \subcaption{}\label{Fig:Data41}
   \end{minipage}

\begin{minipage}{0.25\linewidth}
     \centering
     \includegraphics[scale=0.2]{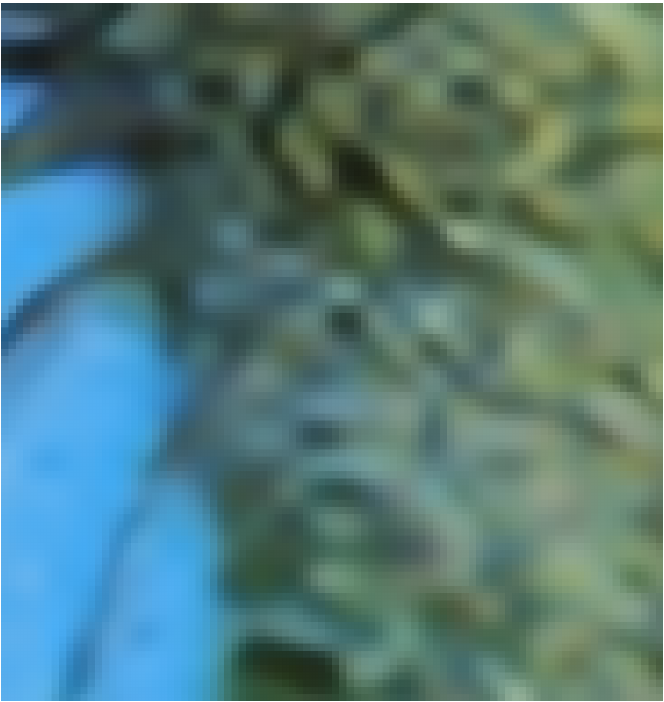}
     \subcaption{}\label{Fig:Data51}
   \end{minipage}\hfill
   \begin{minipage}{0.25\linewidth}
     \centering
     \includegraphics[scale=0.2]{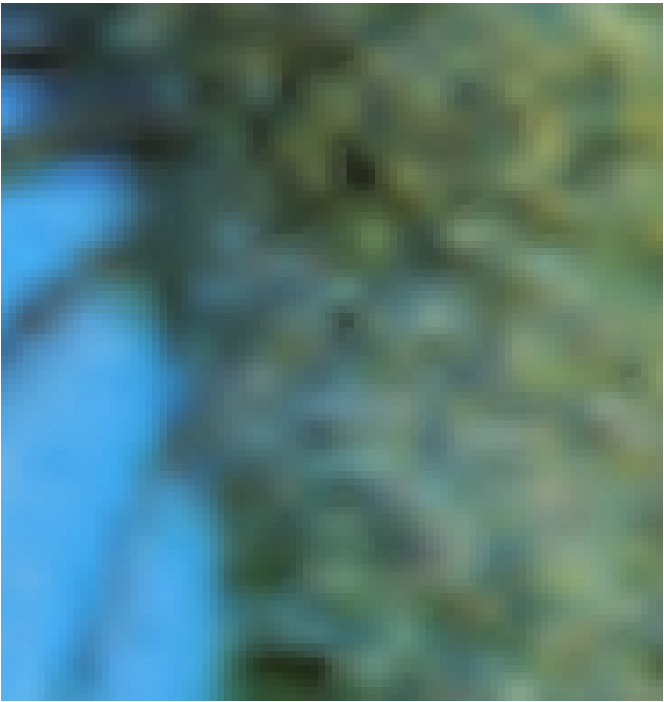}
     \subcaption{}\label{Fig:Data61}
   \end{minipage}\hfill
    \begin{minipage}{0.25\linewidth}
     \centering
     \includegraphics[scale=0.2]{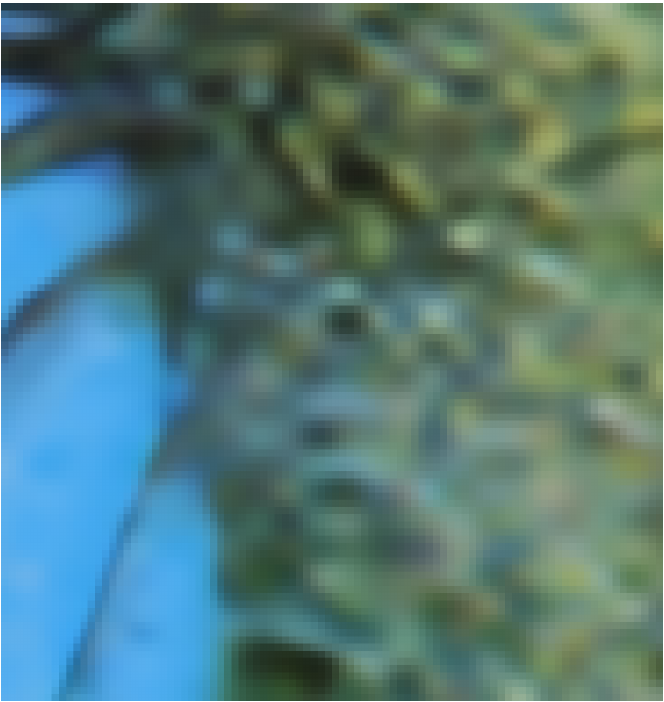}
     \subcaption{}\label{Fig:Data71}
   \end{minipage}\hfill
   \begin{minipage}{0.25\linewidth}
     \centering
     \includegraphics[scale=0.2]{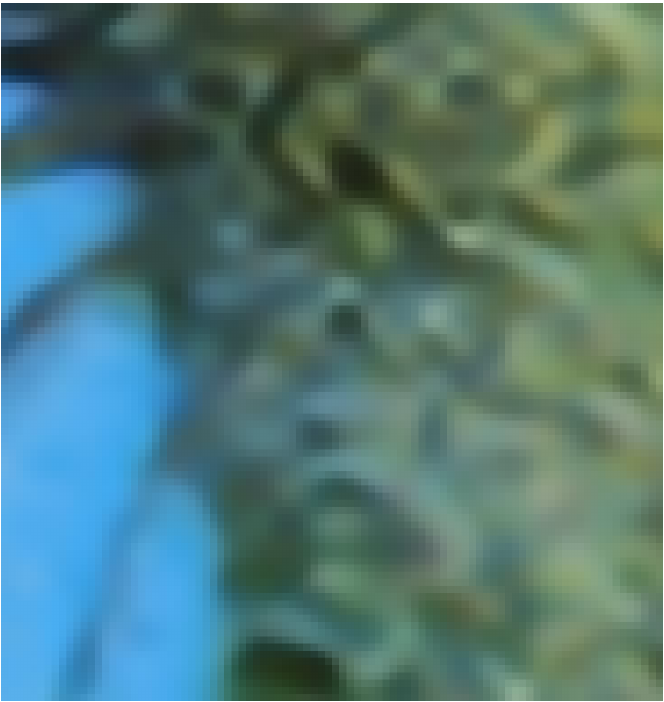}
     \subcaption{}\label{Fig:Data81}
   \end{minipage}
    \end{minipage}
   \end{tabularx}
\caption{Image super resolution with scale $\times$3. (a) Set14:img$\_001$, (b) HR, (c) Bicubic, (d) SRCNN, (e) FSRCNN, (f) LapSRN, (g) VDSR, (h) RDN and (i) SRFRN (ours). }\label{fig6:p6}
\end{figure}
\begin{figure}[!htb]
\begin{tabularx}{\linewidth}{XX}
\begin{minipage}{\linewidth}
     \centering
     \includegraphics[width=0.4\linewidth,height=3cm]{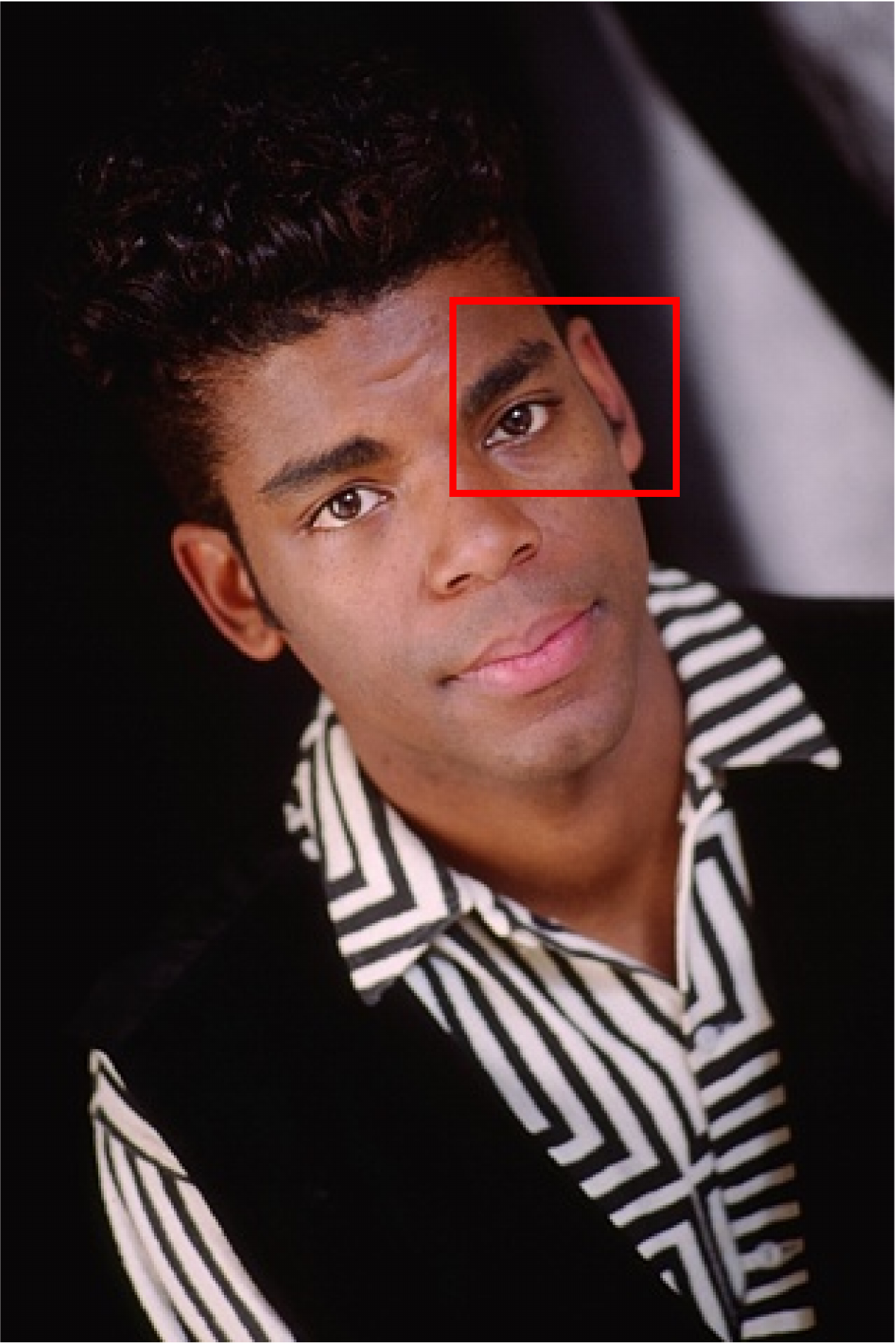}
     \subcaption{}\label{Fig:Data02}
    \end{minipage}
   &
   \begin{minipage}{\linewidth}
   \begin{minipage}{0.25\linewidth}
     \centering
     \includegraphics[scale=0.2]{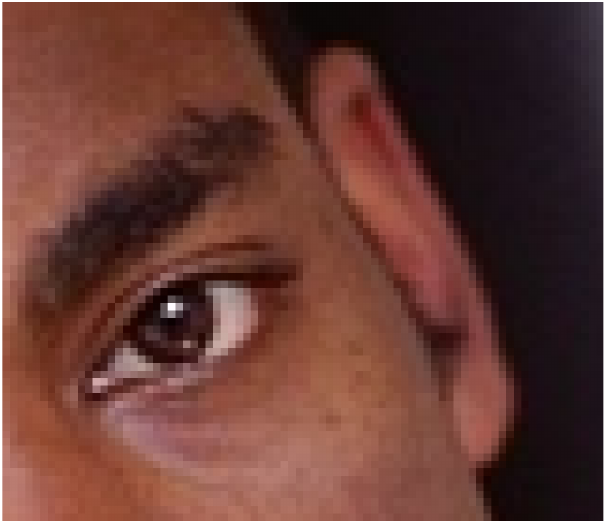}
     \subcaption{}\label{Fig:Data12}
   \end{minipage}\hfill
   \begin{minipage}{0.25\linewidth}
     \centering
     \includegraphics[scale=0.2]{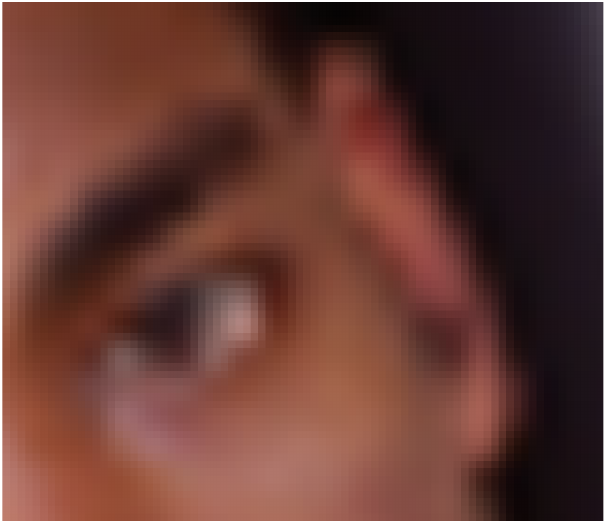}
     \subcaption{}\label{Fig:Data22}
   \end{minipage}\hfill
    \begin{minipage}{0.25\linewidth}
     \centering
     \includegraphics[scale=0.2]{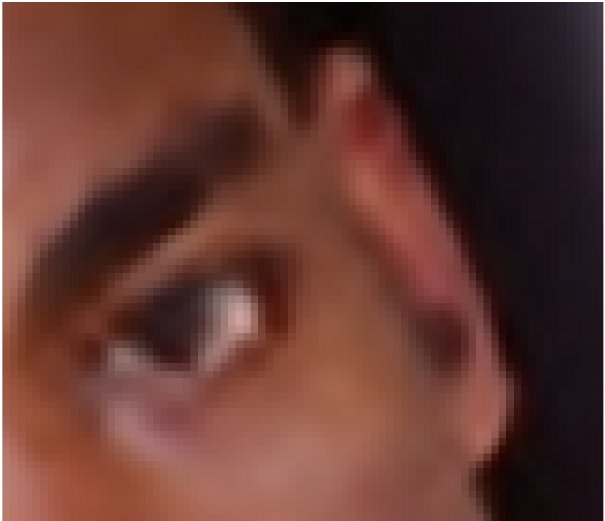}
     \subcaption{}\label{Fig:Data32}
   \end{minipage}\hfill
   \begin{minipage}{0.25\linewidth}
     \centering
     \includegraphics[scale=0.2]{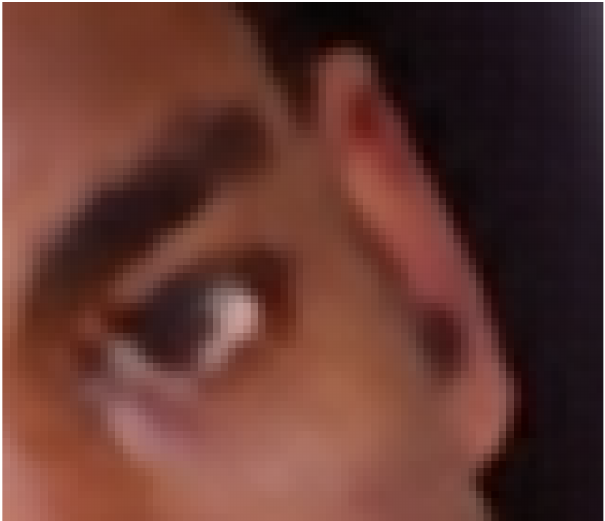}
     \subcaption{}\label{Fig:Data42}
   \end{minipage}

\begin{minipage}{0.25\linewidth}
     \centering
     \includegraphics[scale=0.2]{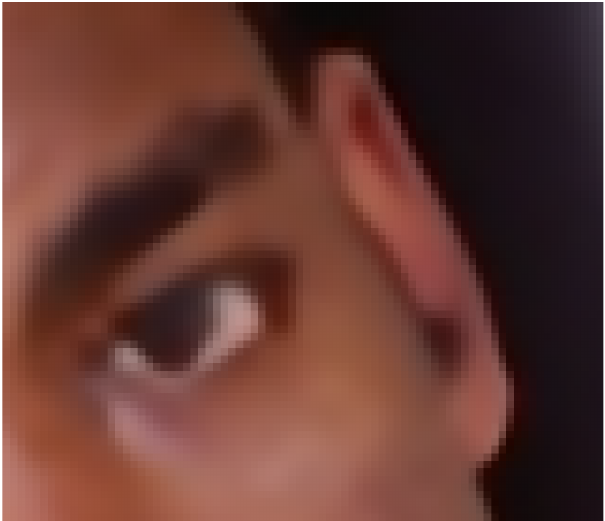}
     \subcaption{}\label{Fig:Data52}
   \end{minipage}\hfill
   \begin{minipage}{0.25\linewidth}
     \centering
     \includegraphics[scale=0.2]{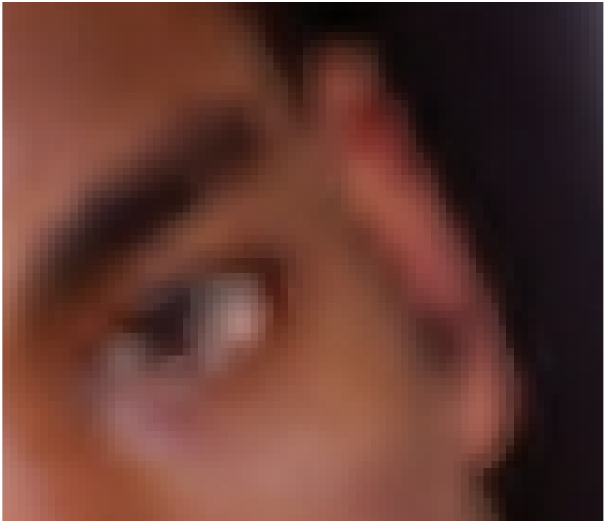}
     \subcaption{}\label{Fig:Data62}
   \end{minipage}\hfill
    \begin{minipage}{0.25\linewidth}
     \centering
     \includegraphics[scale=0.2]{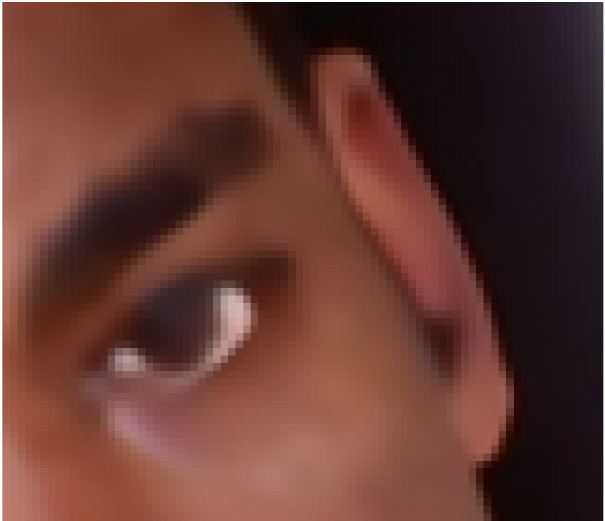}
     \subcaption{}\label{Fig:Data72}
   \end{minipage}\hfill
   \begin{minipage}{0.25\linewidth}
     \centering
     \includegraphics[scale=0.2]{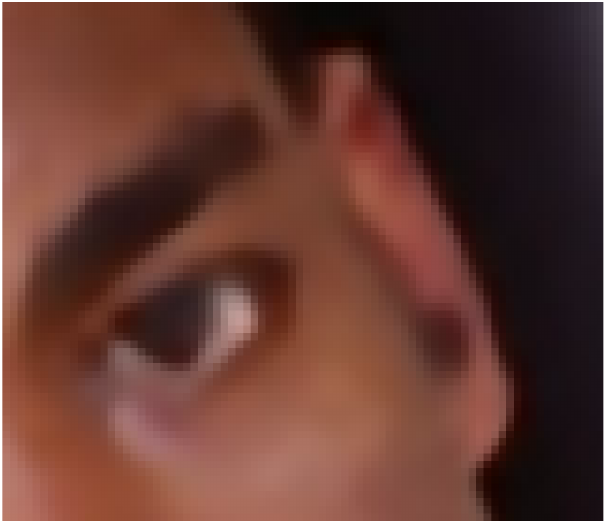}
     \subcaption{}\label{Fig:Data82}
   \end{minipage}
    \end{minipage}
   \end{tabularx}
\caption{Image super resolution with scale $\times$4. (a) BSD100:img$\_063$, (b) HR, (c) Bicubic, (d) SRCNN, (e) FSRCNN, (f) LapSRN, (g) VDSR, (h) RDN and (i) SRFRN (ours). }\label{fig7:p7}
\end{figure}
\begin{figure}[!htb]
\begin{tabularx}{\linewidth}{XX}
\begin{minipage}{\linewidth}
     \centering
     \includegraphics[scale=0.2]{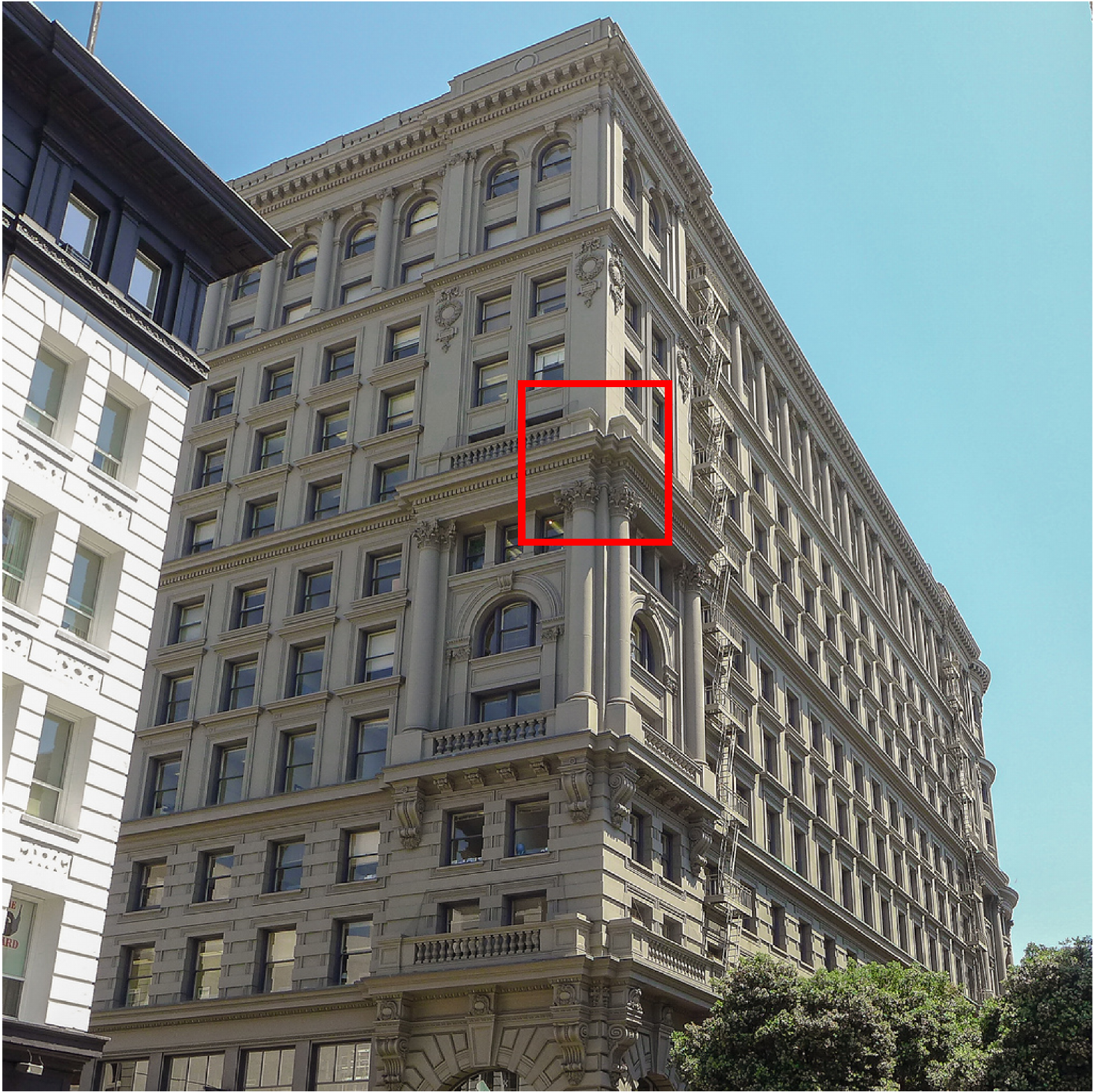}
     \subcaption{}\label{Fig:Data03}
    \end{minipage}
   &
   \begin{minipage}{\linewidth}
   \begin{minipage}{0.25\linewidth}
     \centering
     \includegraphics[scale=0.15]{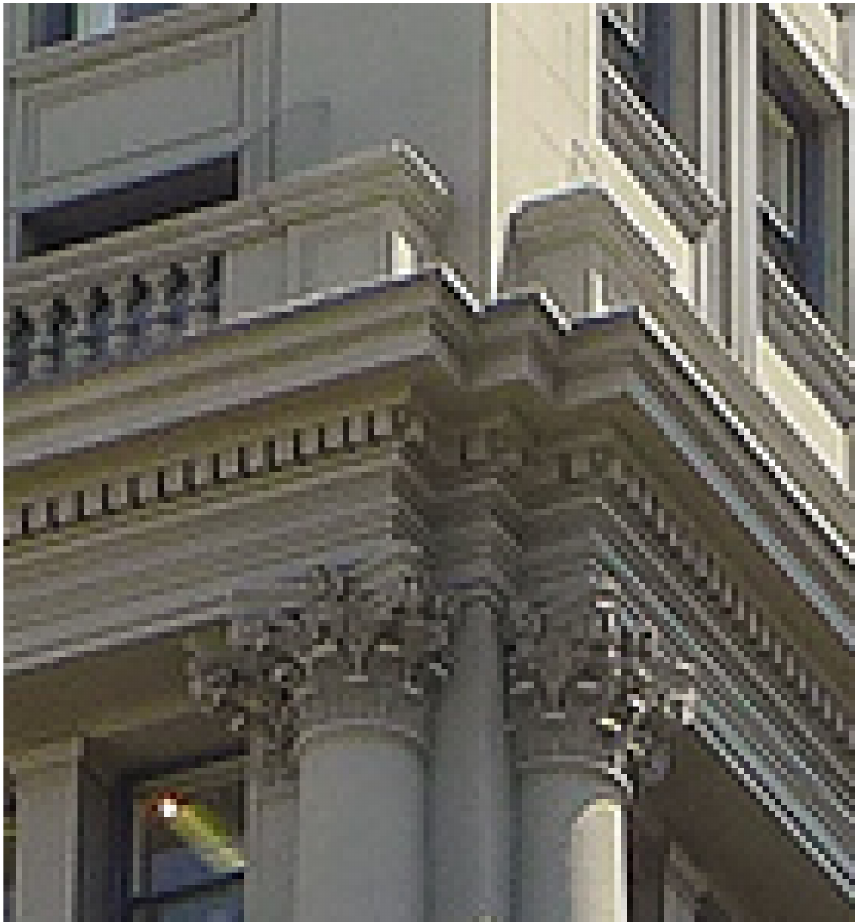}
     \subcaption{}\label{Fig:Data13}
   \end{minipage}\hfill
   \begin{minipage}{0.25\linewidth}
     \centering
     \includegraphics[scale=0.15]{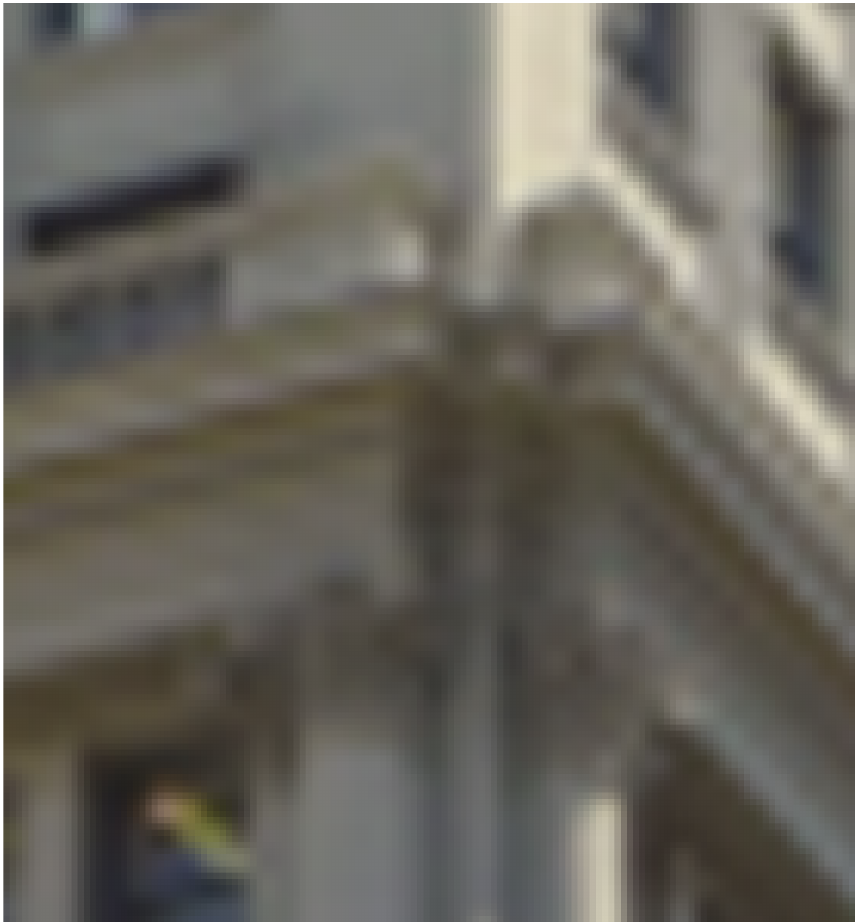}
     \subcaption{}\label{Fig:Data23}
   \end{minipage}\hfill
    \begin{minipage}{0.25\linewidth}
     \centering
     \includegraphics[scale=0.15]{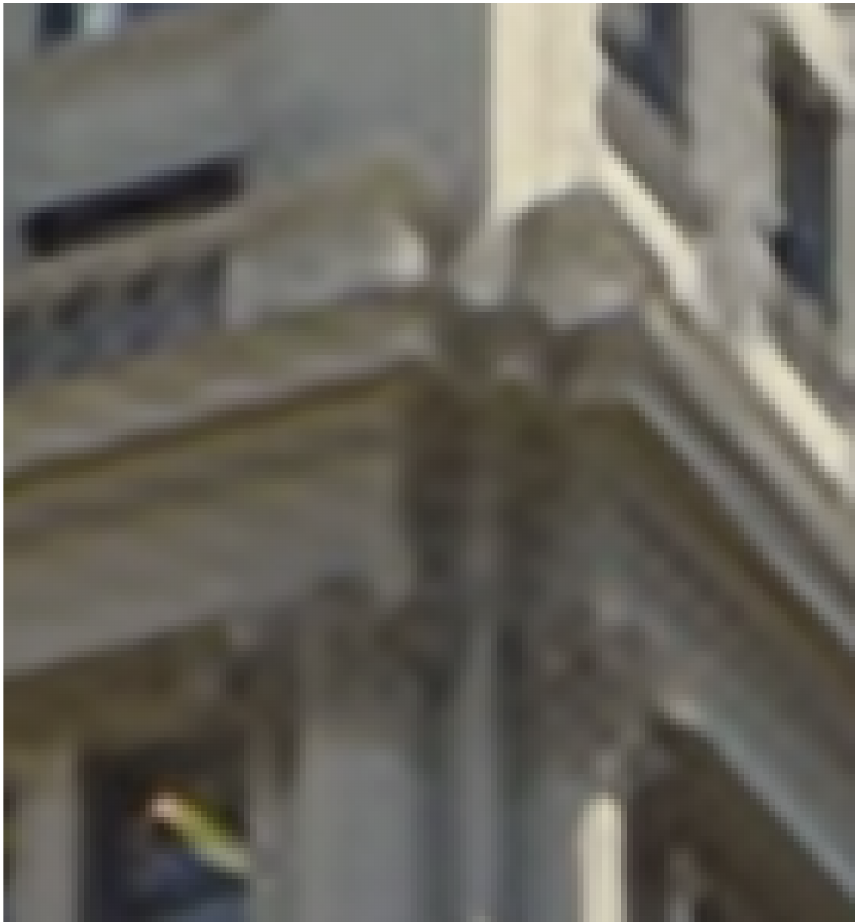}
     \subcaption{}\label{Fig:Data33}
   \end{minipage}\hfill
   \begin{minipage}{0.25\linewidth}
     \centering
     \includegraphics[scale=0.15]{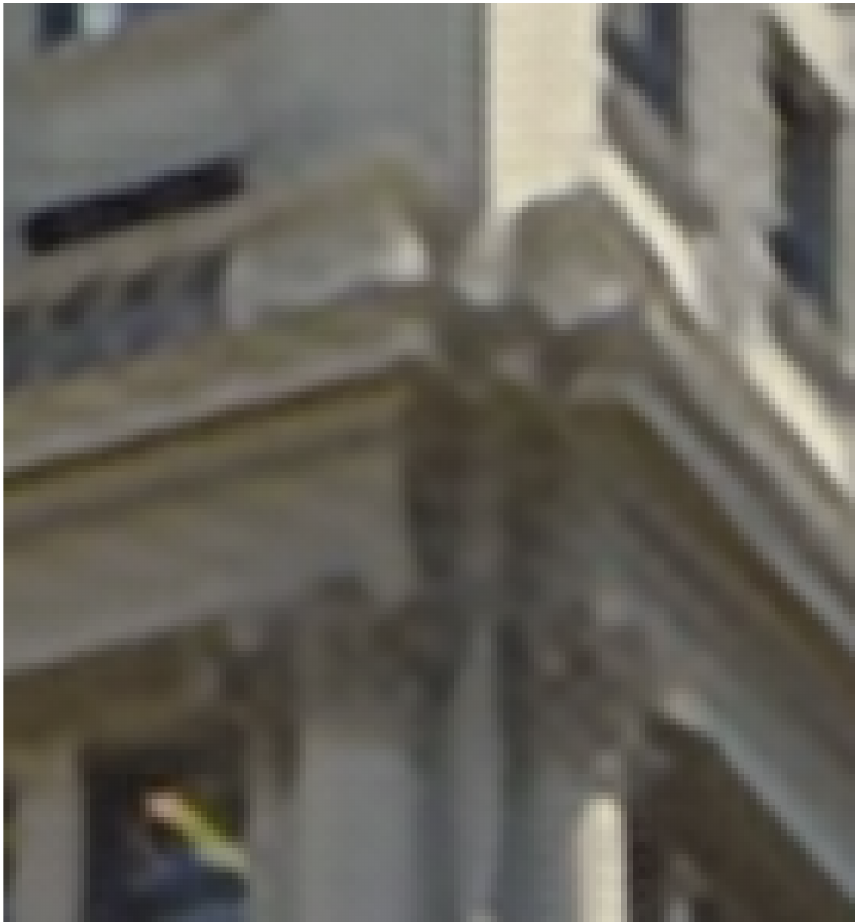}
     \subcaption{}\label{Fig:Data43}
   \end{minipage}

\begin{minipage}{0.25\linewidth}
     \centering
     \includegraphics[scale=0.15]{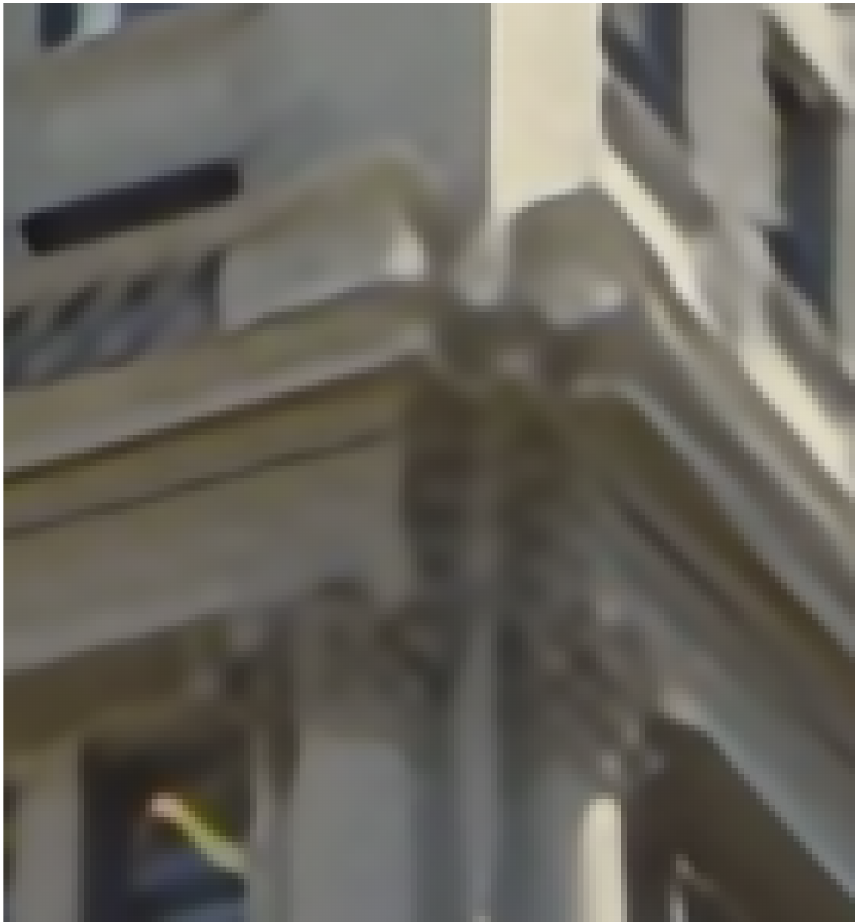}
     \subcaption{}\label{Fig:Data53}
   \end{minipage}\hfill
   \begin{minipage}{0.25\linewidth}
     \centering
     \includegraphics[scale=0.15]{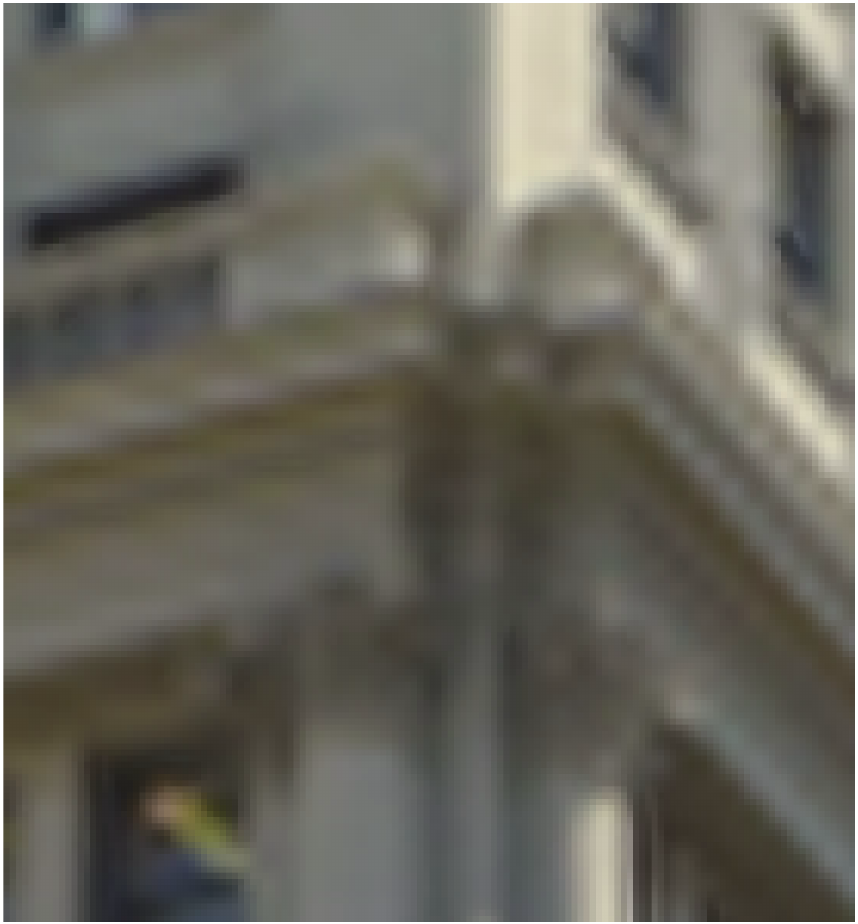}
     \subcaption{}\label{Fig:Data63}
   \end{minipage}\hfill
    \begin{minipage}{0.25\linewidth}
     \centering
     \includegraphics[scale=0.15]{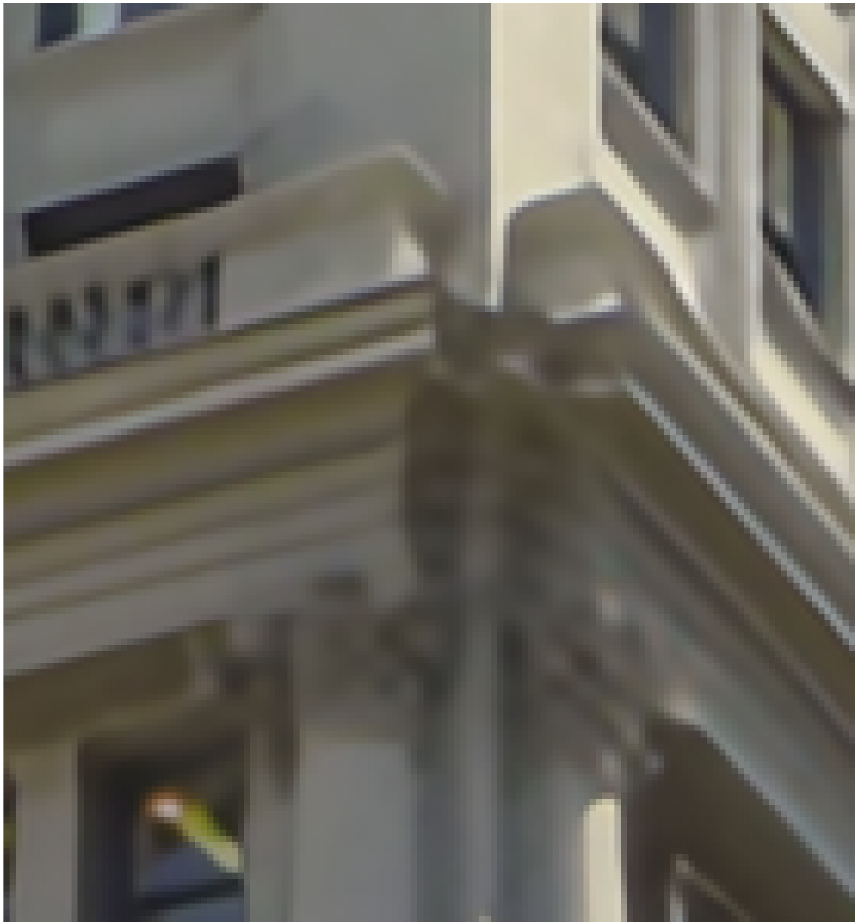}
     \subcaption{}\label{Fig:Data73}
   \end{minipage}\hfill
   \begin{minipage}{0.25\linewidth}
     \centering
     \includegraphics[scale=0.15]{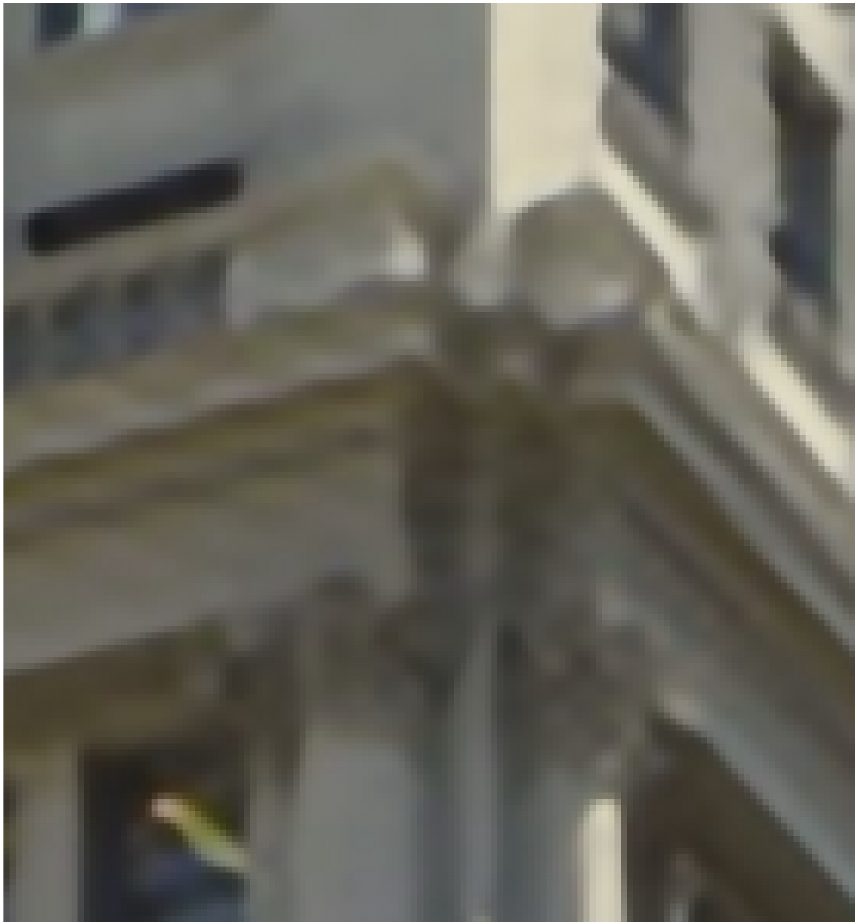}
     \subcaption{}\label{Fig:Data83}
   \end{minipage}
    \end{minipage}
   \end{tabularx}
\caption{Image super resolution with scale $\times$4, (a) Urban100:img$\_014$, (b) HR, (c) Bicubic, (d) SRCNN, (e) FSRCNN, (f) LapSRN, (g) VDSR,  (h) RDN and (i) SRFRN (ours). }\label{fig8:p8}
\end{figure}

\subsection{Comparison with the State-of the Art Methods}
In this section, we provide a comparison of the proposed approach with bicubic interpolation and various other CNN based methods. This quantitative measure is based on two evaluation metrics such as PSNR and SSIM. Based on this, different experiments were conducted on $\times$2, $\times$3 and $\times$4 scales to examine the performance of the proposed algorithm. Table \ref{fig:Table1}  demonstrates the super resolving ability of various approaches such as SRCNN \cite{dong2014learning}, FSRCNN \cite{dong2016accelerating}, VDSR \cite{kim2016accurate}, LapSRN \cite{lai2017deep}, RDN \cite{zhang2018residual}, bicubic interpolation, and our model. The superior performance of the proposed approach is mainly attributed to the residual learning mechanism as well as the shallow nature of the convolutions. Though our approach shows a nominal decrease in PSNR for $\times$2 scales in the case of the Set5 database, the result cannot be generalized since the test database has only 5 images. In contrast, it was observed that for higher scale factors such as $\times$3 and $\times$4, the proposed approach shows a significant rise in PSNR compared to the rest of the state-of-the-art approaches. Further, the most significant increase (4.13 dB) in PSNR for an upscaling factor of 4 can be noted in the Urban100 database with 100 test images. When compared to our approach most of the benchmark algorithms tend to fail at higher scale factors for all the test databases. Besides, Fig. \ref{fig4:p4} shows the comparison of the average PSNR for all the methods in various scales. The traditional CNN based methods exhibit a minimal increase in PSNR compared to each other whereas the proposed method outperforms all of them with a higher margin of PSNR for all the different resolutions.\par
Fig. \ref{fig5:p5}, \ref{fig6:p6}, \ref{fig7:p7} and \ref{fig8:p8} shows the perceptual quality analysis on Set 5, Set 14, BSD100 and Urban100 test databases.  The visual comparison with the benchmark algorithms for scales $\times$3 and $\times$4 are illustrated using these results. Also, the proposed approach has relatively lesser artifacts and obtains clear edges. 
\subsection{Training and Test time Analysis}
\begin{figure*}[h]
  \centering
  \captionsetup{justification=centering}
  \includegraphics[width=1\textwidth, trim={4cm 8cm 4cm 10cm}]{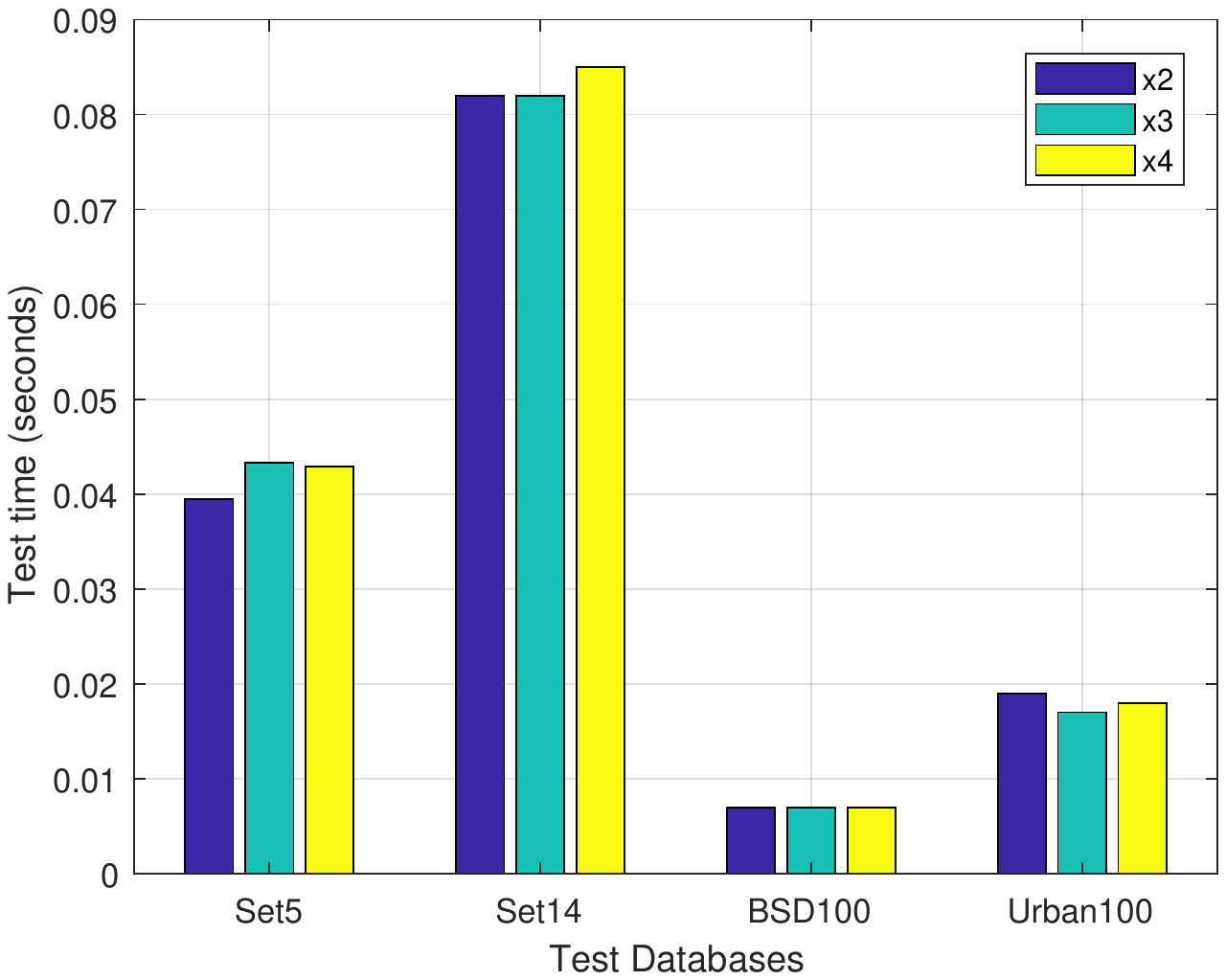}
    \caption{Comparison of test time for different databases and scales for the proposed approach.}
   \label{fig9:p9}
 \end{figure*}

The time and space complexity of the proposed model is evaluated by determining the number of parameters and the test time. Table \ref{fig:Table2} presents the comparison of computational time for an upscaling factor of 4 for the various deep learning methods. Firstly, it is observed that the proposed approach reaches convergence much faster than state-of-the-art methods. In the case of SRCNN, one of the simplest models, it takes 3 days to train even with 12 times lesser parameters compared to the proposed approach. Additionally, with a greater number of parameters, there is a significant increase in training time (1-6 days) for models such as LapSRN, DRRN \cite{8099781}, DRCN \cite{Kim_2016_CVPR} and RDN.  Another model that takes lesser training time is VDSR, but it still has higher test time and low PSNR like SRCNN. In the case of deeper models such as DRRN and DRCN, the test time is more than 1 minute and hence they are impractical to be used in real-time with the hardware available for the experiments. It should be noted that, even though FSRCNN has the least number of learnable parameters, its test time is not significantly different compared to the proposed approach. Hence both approaches have superior time efficiency and cost less memory. But in terms of PSNR and training time, our model achieves better performance than FSRCNN. Hence in overall, the computational complexity of our model is the least compared to the rest of the CNN based approaches.
Fig. \ref{fig9:p9} on the other hand demonstrates the test time for all the databases used in the evaluation of the proposed approach. It can be seen that the test time is comparatively very less and there is only slight variation even when we use higher scales.
\begin{table}[h]
\begin{adjustbox}{width=\textwidth}
\begin{tabular}{|c|c|c|c|c|c|}
\hline
\textbf{Models} & \textbf{PSNR/SSIM (x4)} & \textbf{Train data} & \textbf{Parameters} & \textbf{Test time} & \textbf{Train time} \\ \hline
SRCNN  & 30.48/0.8628 & ImageNet database & 57K   & 0.180 & 3 days    \\
FSRCNN & 30.71/0.8657 & G200 + Yang91     & 12K   & 0.015 & few hours \\
VDSR   & 31.35/0.8838 & G200 + Yang91     & 665K  & 0.120 & 4 hours   \\
LapSRN & 31.54/0.8850 & G200 + Yang91     & 812K  & 0.200 & 3 days    \\
DRRN   & 31.68/0.8888 & G200 + Yang91     & 297K  & 1.210 & 4 days    \\
DRCN   & 31.53/0.8854 & Yang91            & 1.77M & 1.82  & 6 days    \\
RDN   & 32.47/0.8990 & DIV2K            & 22M & 1.56  & 1 day    \\
SRFRN (ours)   & 33.18/0.8488 & G200 + Yang91     & 702K  & 0.041 & 20 min    \\ \hline
\end{tabular}
\end{adjustbox}
\caption{Computational complexity of the different CNN based models.}
    \label{fig:Table2}
\end{table}

 \subsection{PSNR Vs Number of RFR Blocks}
\begin{figure*}[ht]
  \centering
  \captionsetup{justification=centering}
  \includegraphics[width=\textwidth,trim={4cm 8cm 4cm 10cm}]{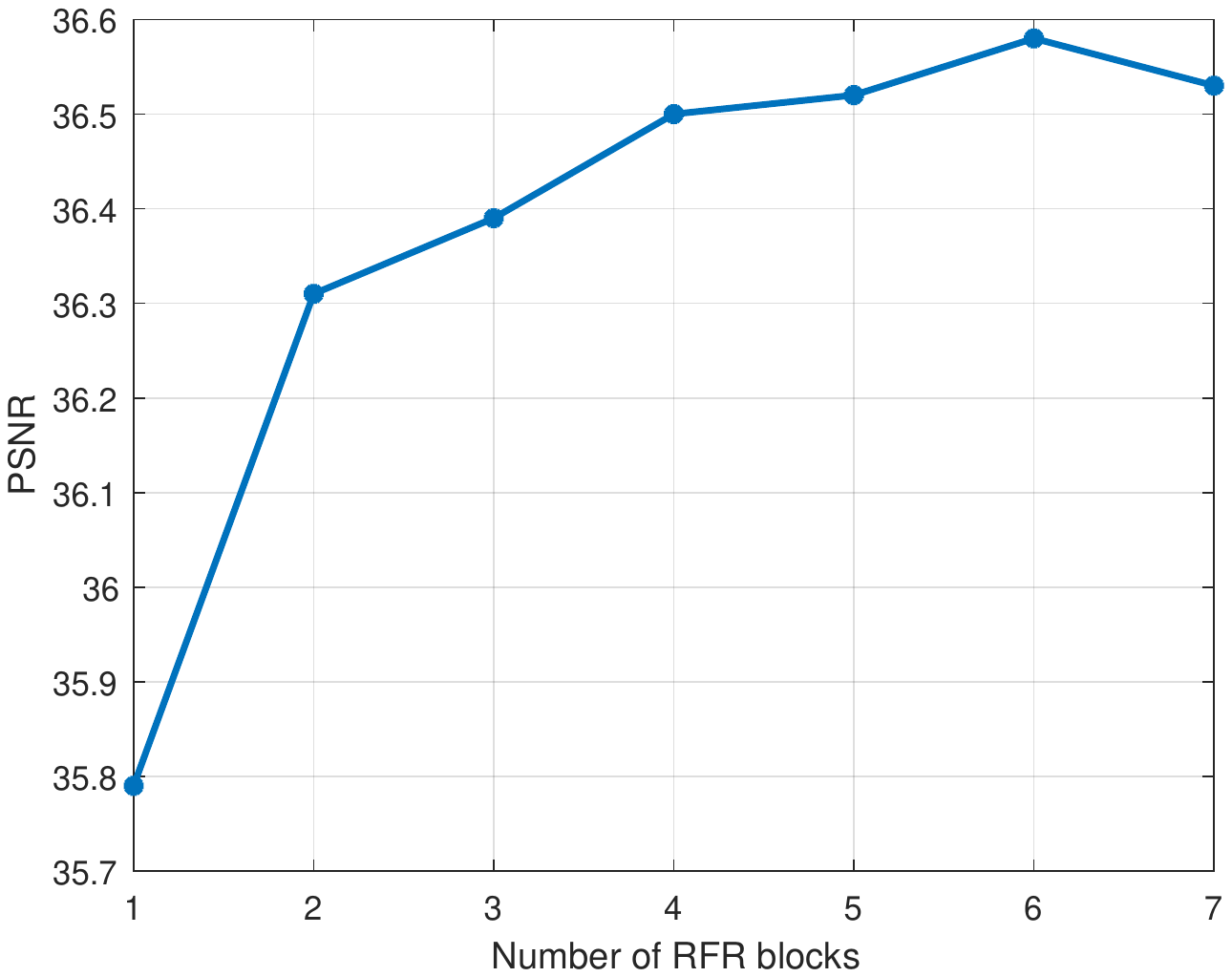}
    \caption{Plot of number of RFR blocks vs PSNR for scale 2.}
   \label{fig10:p10}
 \end{figure*}

 Fig. \ref{fig10:p10} and Table \ref{fig:Table3} illustrates the variation in PSNR with the increase in the number of RFR blocks. In the proposed model, the number of learnable parameters is directly proportional to the number of RFR blocks. The most significant change in PSNR occurs when the number of RFR blocks changes from 1 to 2. Later, the PSNR value in the graph converges by reaching the peak value and then reduces with the increase in the number of RFR blocks. From the table it can be observed that with just 2 RFR blocks the model was able to achieve a PSNR of 36.31, thereby reducing the time complexity. In the following experiments, the highest PSNR of 36.58 was computed using 6 RFR blocks. Thus, the graph illustrates the tradeoff between the number of parameters and PSNR. Finally, the performance of the model drops and the system converges when 7 or more RFR blocks are used.
 
 \begin{table}[ht]
   \begin{adjustbox}{max width=\textwidth}
\begin{tabular}{|l|l|l|l|l|l|l|l|}
\hline
Number of RFR blocks & 1     & 2     & 3     & 4    & 5     & 6     & 7     \\ \hline
Number of parameters & 148K  & 259K  & 370K  & 481K & 592K  & 702K  & 813K  \\ \hline
PSNR (dB)            & 35.79 & 36.31 & 36.39 & 36.5 & 36.52 & 36.58 & 36.53 \\ \hline
\end{tabular}
 \end{adjustbox}
\caption{Variation of PSNR in accordance with number of parameters and RFR blocks.}
    \label{fig:Table3}
   
\end{table}
 
\section{Conclusion}
Super-resolution is a tool of interest in many applications namely, remote sensing, medical applications, entertainment, where these techniques are intended to improve the user experience. In all these applications, besides reconstruction performance, high speed is required, mainly in those applications that demand real-time execution. In this work, we introduced a novel structure for SISR that uses a bicubic interpolated image followed by a set of blocks constructed by a novel structure consisting of convolutional blocks interleaved with feed-forward residual information (RFR units). The bicubic interpolated image is added to the output of the last block to allow the chain of RFR blocks to process differential information. The effect is that the RFR residuals correct the bicubic interpolated image. Further, the training speed is of about 20 minutes, compared to a few hours or days for the state-of-the-art methods, and the test time is on average 41 ms, only worse than the FSRCNN, whose test time is 15 ms. The rest of the methods have a test time 10 to 200 times slower. The reasons for this speed improvement are two. First, the structure uses simpler convolutional blocks, which make the training much faster compared to those structures with deeper and more complex convolutional blocks. Second, the method is based on the construction of residual features, which hold information related only to the interpolation error, not about the image itself, which requires lesser training epochs. Additionally, while testing with $\times$2 scale, the average PSNR of our structure exceeds more than 2dB those of the other methodologies except in the case of RDN. Nevertheless, when the scale is increased, our method shows a lower degradation in PSNR, which is 1.5 dB, compared to the 4 to 5 dB of the average degradation of the rest of the methods. The results show that the accuracy of the introduced architecture is competitive or better than the accuracies shown by the state-of-the-art methods tested in this work.

\section{Acknowledgments}

Authors would like to thank UNM Center for Advanced Research Computing for providing high performance computing, large-scale storage and visualization resources.
 
\bibliography{mybibfile}
\vspace{0.15 in}

\textbf{Meenu Ajith} received the bachelor’s degree in Electronics and Communication Engineering from Amrita school of Engineering in 2015 and the master’s degree in 2017 in Electrical Engineering from The University of New Mexico in 2017. She is currently working towards her PhD degree in Electrical Engineering from The University of New Mexico. Her research interests are Machine Learning, Computer Vision, Pattern Recognition and Image Processing.
\vspace{0.15in}

\textbf{Aswathy Rajendra Kurup} received the bachelor’s degree in Electronics and Communication Engineering from Amrita school of Engineering in 2015 and the master’s degree in Electrical Engineering from The University of New Mexico in 2017. She is currently working towards her PhD degree in Electrical Engineering from The University of New Mexico. Her research interests are Image Processing, Signal Processing and Machine Learning.

\vspace{0.15in}

\textbf{Manel~Mart\'{i}nez~Ram\'on}
is a professor with the ECE department of The University of New Mexico. He holds the King Felipe VI Endowed Chair of the University of New Mexico, a chair sponsored by the Household of the King of Spain. He is a Telecommunications Engineer (Universitat Politecnica de Catalunya, Spain, 1996) and PhD in Communications Technologies (Universidad Carlos III de Madrid, Spain, 1999). His research interests are in Machine Learning applications to smart antennas, neuroimage, first responders and other cyber-human systems, smart grid and others. His last work is the monographic book “Signal Processing with Kernel Methods”, Wiley, 2018.

\end{document}